\documentclass[aps,pra,twocolumn,floatfix,longbibliography]{revtex4-1}
\usepackage{graphicx}
\usepackage{amsmath}
\usepackage{braket}
\usepackage{tikz}
\usepackage{comment}
\begin{document}

\title{Hyperspherical approach to dipolar Bose-Einstein condensates beyond the mean-field limit}
\begin{abstract}
We apply a hyperspherical formulation to a trapped Bose-Einstein condensate with dipolar and contact interactions. Central to this approach is a general correspondence between K-harmonic hyperspherical methods and a suitable Gaussian {\it ansatz} to the Gross-Pitaevskii equation, regardless of the form of the interparticle potential. This correspondence allows one to obtain hyperspherical potential energies for a wide variety of physical problems. In the case of the dipolar Bose-Einstein condensate, this motivates the inclusion of a beyond-mean field term within the hyperspherical picture, which allows us to describe the energies and wavefunctions of excitations of self-bound dipolar droplets outside of the mean-field limit.

\end{abstract}

\author{Eli J. Halperin}
\affiliation{JILA, NIST, and Department of Physics, University of Colorado, Boulder, Colorado 80309-0440, USA}
\author{John L. Bohn}
\affiliation{JILA, NIST, and Department of Physics, University of Colorado, Boulder, Colorado 80309-0440, USA}
\date{\today}
\maketitle

\section{Introduction}

Recently a number of interesting phenomena have been observed in dipolar Bose-Einstein Condensates (BEC), including dipolar droplets~\cite{ferrier2016observation,kadau2016observing,chomaz2016quantum,schmitt2016self}. Dipolar droplets are self-bound collections of strongly dipolar atoms, where the droplets are elongated along the polarization axis. Here interatomic attraction is balanced by quantum fluctuations, leading to a metastable state which slowly evaporates due to three-body recombination.  Dipolar droplets are considered self-bound as the droplet remains intact even after the trap is entirely removed.

Upon their discovery, it was immediately realized that these dipolar droplets could be described theoretically by an extended Gross-Pitaevskii equation (EGPE), the extension being the inclusion of fluctuation terms beyond the mean-field description \cite{lima2011quantum}.  The EGPE describes the self-binding and stability of the droplet \cite{baillie2016droplet,bisset2016ground}, as well as its low-lying excitations \cite{lima2011quantum,baillie2017collective}. 

In this paper we present an alternative theoretical approach to dipolar droplets, based on an explicit wavefunction-based method. The method relies on choosing a small set of collective coordinates that represent the spatial extent of the full gas of atoms, rather than the coordinates of any one atom.  By this means the basic properties of the gas are described, in the examples herein, by a two-dimensional, {\it linear} Shcr\"{o}dinger equation.  The usual intuitions of quantum mechanics can be applied, and explicit, albeit approximate, wave functions for excitations of the BEC can be shown.  The method has similarities with the variational Gaussian {\it ansatz} approach for finding approximate ground state solutions to the EGPE \cite{baillie2016droplet,bisset2016ground,wachtler2016ground}, as we will explore in detail.  However, our method has the advantage that excited states of collective motion can be calculated as well.  

The method re-casts collective radial and axial coordinates of the atoms into their root-mean-square average coordinates, averaged over all atoms.  These coordinates encompass the most basic properties of the BEC, namely, their overall size and collective excitations in radial and axial degrees of freedom.  Mathematically, they represent  hyperradii in a multidimensional configuration space, whereby the method is referred to as a hyperspherical approach.

Hyperspherical approaches to BEC have proven fruitful in the past, describing, for example, the stability of BECs with attractive contact potentials \cite{bohn1998effective,watson1999improved,biswas2010stability}; condensate fraction \cite{haldar2014condensate}; multicomponent BECs \cite{sogo2004stability}; the influence of realistic two body contact interactions, including effective range corrections \cite{sorensen2002two,sorensen2003correlated,sorensen2003structure} and even formally infinite scattering lengths \cite{thogersen2007trapped, lekala2014behavior,ding2017renormalized,sze2018hyperspherical}; realistic two-body interactions \cite{das2004potential,das2007behavior,chakrabarti2008shape}; and condensate dynamics \cite{van2000collapse,sogo2005coherent,liu2007time}.  These treatments are all necessarily approximate, yet an exciting recent development shows that their accuracy can be enhanced by combining hyperspherical coordinates with solutions to the Gross-Pitaevskii equation (GPE)~\cite{PhysRevA.103.023325}.  Further, hyperspherical methods can be fruitfully applied to fermionic gases as well~\cite{rittenhouse2006hyperspherical,rittenhouse2008degenerate,rittenhouse2009collective}.

This paper extends the hyperspherical approach to the case of a dipolar BEC for the first time. Instead of using a single hyperradius to describe the Bose gas, we use two hyperradii~\cite{kushibe2004aspects}, describing the average displacement of the particles in the radial and axial directions. This allows us to effectively incorporate the dipole-dipole interaction into hyperspherical coordinates. In our approximation the two hyperradii are the only coordinates. We ignore the explicit dependence of the wave function on hyperangles, an approximation known as the K-harmonic approximation.  This approach is therefore aimed at describing the ground state of the condensate, as well as collective excitations such as breathing and quadrupole modes, or variations of these.

Significantly, we show a general correspondence between all K-harmonic hyperspherical approaches and the Gaussian {\it ansatz} to the GPE regardless of the interparticle potential. We see that, in the limit of a large number of particles, the effective potential in the hyperspherical 1 dimensional (1D) or 2 dimensional (2D) Schr\"odinger equation approaches the energy surface given by a suitable 1D or 2D Gaussian {\it ansatz} to the GPE. This insight vastly simplifies an entire class of hyperspherical calculations when one is concerned with many-body physics. In the hyperspherical method, this energy surface plays the role of the potential energy, and thus we can compute excited states and condensate dynamics within the 2D Schr\"odinger equation, whereas the variational Gaussian {\it ansatz} is limited to ground states and low-lying excitations in a harmonic approximation. One can proceed alternatively by considering Bogoliubov excitations occurring on top of a variational ground state~\cite{blakie2020variational,hu2020collective}, a topic we will not pursue here.

Section \ref{hyper_background} outlines the hyperspherical method with two hyperradii, showing the effective 2D Schr\"odinger equation. Section \ref{GPE_corr} shows the more general correspondence between the K-harmonic hyperspherical approximation and the Gaussian {\it ansatz} to the GPE. This allows us to translate the Lee, Huang, and Yang (LHY) correction~\cite{LHY1,LHY2} into hyperspherical coordinates. We then apply this approach to the self-bound dipolar droplet in section~\ref{sec:droplet}, showing excited states and spectra of the dipolar BEC. 

\section{The hyperspherical approach}
\label{hyper_background}

We consider a collection of $N$ identical dipolar atoms of mass $m$ in a cylindrically symmetric harmonic potential $V_{\text{ext}}$ with radial trapping frequency $\omega_\rho$ and axial trapping frequency $\omega_z$. The Hamiltonian is

\begin{align}
H =  \sum_{i=1}^N \left[ - \frac{ \hbar^2 }{ 2m } \nabla_i^2 + V_{\text{ext}} ({\bf r}_{i})  \right] + \sum_{i<j}^N \left[ V_{dd}({\bf r}_{ij}) +  V_c({\bf r}_{ij}) \right] ,
\label{full_ham}
\end{align}
where $V_\text{ext}$ is the external trapping potential, and the dipole-dipole interaction, $V_{dd}$, pertains to a pair of dipoles polarized along the laboratory $z$ axis
\begin{align}
V_{dd}({\bf r_{ij}}) =  \frac{3\hbar^2}{m} \frac{a_{dd}}{r^3} (1-3\cos^2\theta).
\end{align}
Here the dipole length is defined as $a_{dd} \equiv m\mu_0\mu^2/12\pi\hbar^2$, and $\theta$ is the angle between $\bf{r_{ij}}$ and the polarization axis. The two-body contact potential is given by
\begin{align}
    V_c({\bf r_{ij}}) = \frac{4\pi \hbar^2 a}{m} \delta ({\bf r_{ij}}),
\end{align}

Given that the dipole-dipole interaction obeys a cylindrical but not spherical symmetry (i.e. no $\varphi$ dependence) and the other terms in the Hamiltonian have full spherical symmetry, we wish to describe a BEC only in terms of it's height in the $z$ direction and it's width in the $x$-$y$ plane. To this end we introduce two collective hyperradii via
\begin{eqnarray}
P^2 &=& \frac{ 1 }{ N } \sum_{i=1}^N (x_i^2 + y_i^2 ), \\
Z^2 &=& \frac{ 1 }{ N } \sum_{i=1}^N z_i^2 .
\end{eqnarray}
Here $P$ (understood as capital $\rho$) gives the root-mean-square displacement of the gas in the radial direction and $Z$ the root-mean-square displacement in the axial direction. These two coordinates, along with $3N - 2$ additional hyperangles, describe the complete configuration of the gas. In these coordinates, the external harmonic potential $V_\text{ext}$ has a simple form
\begin{align}
    V_\text{ext} &= \sum_i^N \left[ \frac{1}{2} m \omega_\rho \rho_{i}^2 + \frac{1}{2} m \omega_z z_i^2 \right], \\
     &=  \frac{1}{2} M \omega_\rho P^2 + \frac{1}{2} M \omega_z Z^2,
\end{align}
where $M =m N$. The kinetic term $T$ in eq.~\eqref{full_ham} can now be written as~\cite{bohn1998effective,Smirnov77}
\begin{align}
T&=- \frac{ \hbar^2 }{ 2M } \left[ \frac{ 1 }{ P^{2N-1} } \frac{ \partial }{ \partial P } \left( P^{2N-1} \frac{ \partial }{ \partial P } \right)  - \frac{  \Lambda_P^2 }{ P^2 } \right] \nonumber \\
&- \frac{ \hbar^2 }{ 2M } \left[ \frac{ 1 }{ Z^{N-1} } \frac{ \partial }{ \partial Z} \left( Z^{N-1} \frac{ \partial }{ \partial Z } \right) - \frac{  \Lambda_Z^2 }{ Z^2 } \right].
\end{align}

Here $\Lambda_P$ and $\Lambda_Z$ are the two grand angular momentum operators, and behave in analogy with the 3D angular momentum operator. They are given by
\begin{align}
\Lambda_P^2 &= \sum_{i<j} \left[ x_i \frac{\partial}{\partial x_j}  - x_j \frac{\partial}{\partial x_i}\right]^2, \\ 
\Lambda_Z^2 &= \sum_{i<j} \left[ z_i \frac{\partial}{\partial z_j}  - z_j \frac{\partial}{\partial z_i}\right]^2.
\end{align}
Where in the first sum, $x_i$ is understood to be an element of the $2N$ dimensional vector $(x_1,y_1,x_2,y_2, \cdots, x_N,y_N)$. These operators obey the eigenvalue equations~\cite{Smirnov77,avery1997fourier}
\begin{align}
    \Lambda_P Y_{\lambda \mu}^P &= \lambda ( \lambda + 2N -2) Y_{\lambda \mu}^P, \\
        \Lambda_Z Y_{\lambda \mu}^Z &= \lambda ( \lambda + N -2) Y_{\lambda \mu}^Z.
\end{align}
Here $\mu$ stands for a degenerate set of indices for each eigenvalue $\lambda$ that we will not specify here. These eigenvectors $Y_{\lambda \mu}^{(P,Z)}$ are the hyperspherical harmonics. $Y_{\lambda \mu}^P$ forms a complete orthonormal basis for $2N - 1$ hyperangles and $Y_{\lambda \mu}^Z$ for $N-1$ hyperangles.  It is implied that they are symmetric under exchange of identical bosons, although we need not perform this symmetrization explicitly for our purposes. We then expand an arbitrary many-body wavefunction $\psi$ as
\begin{align}
\psi = P^{(2N-1)/2} Z^{(N-1)/2} \sum_{\lambda \lambda' \mu \mu'} F_{\lambda \lambda' \mu \mu'}(P,Z) Y_{\lambda \mu}^P Y_{\lambda' \mu'}^Z.
\label{eq:psi_expansion}
\end{align}
The prefactor of $P^{(2N-1)/2} Z^{(N-1)/2}$ eliminates any first derivatives in $T \psi$. We end up with a new set of coupled Schr\"odinger equations,
\begin{widetext}
\begin{align}
\Bigg(& -\frac{ \hbar^2 }{ 2M } \left[ \frac{ \partial^2 }{ \partial P^2 } + \frac{ \partial^2 }{ \partial Z^2 }  - \frac{(2N -1)(2N-3) +  4\lambda(\lambda +2N -2) }{ 4 P^2 }
 - \frac{(N -1)(N-3) +  4\lambda'(\lambda' +N -2) }{ 4Z^2 } \right]+ \frac{1}{2} M \omega_\rho P^2  \nonumber \\
  &+ \frac{1}{2} M \omega_z Z^2 \Bigg) F_{\lambda \lambda' \mu \mu' } +  \sum_{\bar{\lambda} \bar{\lambda}' \bar{\mu} \bar{\mu}' } \left[ \sum_{i<j} \bra{\bar{\lambda} \bar{\lambda}' \bar{\mu} \bar{\mu}' } V_c (\mathbf{r_{ij}}) + V_{dd} (\mathbf{r_{ij}}) \ket{\lambda \lambda' \mu \mu'} \right] F_{\bar{\lambda} \bar{\lambda}' \bar{\mu} \bar{\mu}' }(P,Z) = E F_{\lambda \lambda' \mu \mu'}.
\end{align}
\end{widetext}
We expect that the general features of the condensate will emerge with a rather small expansion of states~\cite{bohn1998effective}. In fact, we choose the smallest possible expansion, known as the K-harmonic approximation, and thus set $\lambda = \lambda^{\prime}=0$.  We suppress the notation $ \lambda \lambda^{\prime} \mu \mu^{\prime} $ in the following and denote the lowest hyperspherical harmonic as $|0\rangle$. We expect this should approximately represent the ground state and bulk dynamics of the condensate. Making this approximation, we have the simplified equation
\begin{widetext}
\begin{align}
    \bigg(& - \frac{ \hbar^2 }{ 2M } \left[ \frac{ \partial^2 }{ \partial P^2 } + \frac{ \partial^2 }{ \partial Z^2 }
- \frac{ (2N-1)(2N-3) }{ 4 P^2 } - \frac{ (N-1)(N-3) }{ 4 Z^2 } \right] + \frac{1}{2} M \omega_\rho^2 P^2 + \frac{1}{2 } M \omega_Z^2 Z^2 \nonumber \\ &+ \sum_{i<j}  \bra{0} V_c (\mathbf{r_{ij}}) + V_{dd} (\mathbf{r_{ij}})\ket{0} \bigg) F_0(P,Z) = E F_0(P,Z).
\label{2D_schrodinger}
\end{align}
\end{widetext}

\section{General relationship to the Gross-Pitaevskii equation}
\label{GPE_corr}
We first consider the more general case where the two-body potential $V(\mathbf{r_{ij}})$ depends on the specific vector $\mathbf{r_{ij}}$ between two atoms. This will help us compute the specifics cases of $V_c$ and $V_{dd}$. Selecting the single pair of particles $i=1$, $j=2$, we define the hyperangles $\alpha$ and $\beta$ such that
\begin{align}
    \rho_{12} &= \sqrt{2 N} P \sin \alpha,
    \label{eq:hyp_coors1} \\
    z_{12} &= \sqrt{2 N} Z \sin \beta.
    \label{eq:hyp_coors2}
\end{align}
These hyperangles have associated hyperspherical surface area elements~\cite{Smirnov77}
\begin{align}
d \Omega^\alpha &= \sin \alpha \cos^{2N - 3} \alpha~d \alpha, \\
d \Omega^\beta &= \cos^{N-2} \beta~d \beta
\end{align}
Let $\phi$ give the angle for the unit vector $\hat \rho_{12}$. Then the total surface area over the entire hypersphere is
\begin{align}
    d \Omega^N d \Omega^{2N} &=  \sin \alpha \cos^{2N - 3} \alpha \cos^{N-2} \beta ~d \alpha d \phi d \beta \nonumber \\
    &\times d\Omega^{N-1} d \Omega^{2N-2}.
\end{align}
The hyperspherical harmonics $Y_0^P$ and $Y_0^Z$ are constant across their respective hyperspheres. Since they are normalized, we have
\begin{align}
    Y_0^P = \sqrt{\frac{\Gamma(N)}{2 \pi^{N}}} =  1 / \sqrt{I(2N)}, \\
    Y_0^Z = \sqrt{\frac{\Gamma(N/2)}{2 \pi^{N/2}}} =  1 / \sqrt{I(N)}.
\end{align}
Where $I(k)$ stands for the hyperspherical surface area in $k$ dimensions. Since these harmonics do no depend on hyperangles, each term in the sum in eq.~\eqref{2D_schrodinger} is the same. Thus we have
\begin{align}
    V_\text{int} &= \sum_{i<j}  \bra{0} V(\mathbf{r_{ij}}) \ket{0}, \\
    &=  \frac{N(N-1)}{2} \bra{0} V(\mathbf{r_{12}}) \ket{0}.
\end{align}
Using our definitions of the hyperangles, 
\begin{widetext}
\begin{align}
V_{\text{int}} &= \frac{N(N-1)}{2} \frac{1}{I(2N) I(N) } \int V(\mathbf{r_{12}}) \sin \alpha \cos^{2N - 3} \alpha \cos^{N-2} \beta ~d \alpha d \phi d \beta \int d\Omega^{N-1} \int d \Omega^{2N-2} \\ 
&= \frac{N(N-1)I(N-1) I(2N -2)}{2 I(2N) I(N) (2N)^{3/2} P^2 Z}\int V(\mathbf{r_{12}}) \rho_{12} \left( 1 - \frac{z_{12}^2}{2N Z^2} \right)^{(N - 3)/2}  \left( 1 - \frac{\rho_{12}^2}{2N P^2}\right)^{N-2} ~d \rho_{12} d z_{12} d \phi \label{exact_2d_int} \\
&\approx \frac{N^2}{8 \pi^{3/2}} \frac{1}{P^2 Z} \int V(\mathbf{r_{12}})  \exp(-z_{12}^2 / 4 Z^2 )  \exp(-\rho_{12}^2 / 2 P^2)\rho_{12}~d \rho_{12} d z_{12} d \phi.
\label{largeN_Vint}
\end{align}
\end{widetext}
the first line rephrases the angles in terms of interparticle coordinates according to eqs.~(\ref{eq:hyp_coors1},~\ref{eq:hyp_coors2}). In the last line we went to the large $N$ limit, using that
\begin{align}
    e^{x}=\lim _{n \rightarrow \infty}\left(1+\frac{x}{n}\right)^{n},
\end{align}
and that $N \approx N-2 \approx N-3$. We also simplified the ratio of Gamma functions present in the hyperspherical surface area elements. This expression is exact until the final line, where Gaussians emerge from the hyperangular volume elements. 

Now we will compare this expression to the Gaussian {\it ansatz} to the GPE. The GPE, for purely contact interactions, is given by~\cite{dalfovo1999theory}
\begin{align}
i\hbar\frac{\partial\phi}{\partial t} = \left[-\frac{\hbar^2}{2m} \frac{\partial^2}{\partial x^2} + V_\text{ext} + g\vert\phi\vert^2 \right] \phi.
\label{GPE}
\end{align}
Here $g = 4 \pi \hbar^2 a /m$, with $a$ the scattering length. One can find approximate ground states to eq.~\ref{GPE} using a variational wavefunction and then minimizing the resulting energy functional. The 2D Gaussian {\it ansatz} to the GPE posits a variational wavefunction of the form
\begin{align}
\phi = \left( \frac{N}{\pi^{3/2} \sigma_\rho^2 \sigma_z}\right)^{1/2} \exp \left( \rho^2 / 2 \sigma_\rho^2 \right) \exp \left( z^2 / 2 \sigma_z^2 \right).
\label{ansatz_2dgpe}
\end{align}
This Gaussian ansatz wave function $\phi$ can be transformed to hyperspherical coordinates, giving the correspondence
\begin{align}
     Z &= \sigma_z / \sqrt{2}, 
    \label{eq:correspodance1}\\
     P &= \sigma_\rho.
     \label{eq:correspodance2}
\end{align}

In the variational {\it ansatz} to the GPE, the energy due to an arbitrary interaction is
\begin{align}
E_{\text{int}} = \frac{1}{2} \int d \mathbf{r_1} d \mathbf{r_2} |\phi_1|^2 |\phi_2|^2 V(\mathbf{r_{12}}).
\end{align}
Making a change of coordinates and computing this integral, we find that
\begin{align}
E_{\text{int}} &= \frac{N^2}{4 \sqrt{2} \pi^{3/2}} \frac{1}{\sigma_\rho^2 \sigma_z}\int V(\mathbf{r_{12}}) \exp(-z_{12}^2 / 2 \sigma_z^2 ) \nonumber \\
&\times \exp(-\rho_{12}^2 / 2 \sigma_\rho^2)\rho_{12}~d \rho_{12} d z_{12} d \phi.
\label{gaussian_gpe_simple}
\end{align}
 By using the correspondence given in eqs.~(\ref{eq:correspodance1},~\ref{eq:correspodance2}), eq.~\eqref{gaussian_gpe_simple} is seen to match eq.~\eqref{largeN_Vint}, i.e. $V_\text{int} = E_\text{int}$. Other authors have noted the seeming similarity between these hyperspherical potentials $V_\text{int}$ and energy functional of the Gaussian {\it ansatz} to the GPE, c.f. Ref.~\cite{PhysRevA.103.023325}. We have shown that this resemblance is not simply coincidental, in the limit of a large number of particles they are in fact the same. The convergence in $N$ is rapid for typical potentials; for $N=10,000$ particles and a delta function interaction, the hyperspherical potential matches the Gaussian {\it ansatz} energy surface to within $0.03 \%$. Additional terms in the Gaussian {\it ansatz} energy surface given by the external potential and the kinetic energy will match the external potential and centrifugal terms in the hyperspherical picture, respectively. In the 1D case, with a single hyperradius and a spherically symmetric Gaussian {\it ansatz}, the two approaches can also be shown to give the same potential energies in the large $N$ limit, following a similar argument as above. However, the hyperspherical approach allows one to go beyond the Gaussian {\it ansatz} to the GPE, due to the kinetic energy terms in $P$ and $Z$, which have no counterpart in the Gaussian {\it ansatz} approach. These terms allow us to compute excited states of the condensate and to extract a spectrum. 

\section{Dipolar Droplet}
\label{sec:droplet}
In order to adequately describe the dipolar droplet, it is necessary to incorporate the effects of fluctuations that go beyond the mean-field~\cite{baillie2016droplet,PhysRevA.94.033619}. We can exploit the close connection between our K-harmonic theory and the Gaussian {\it ansatz} to proceed as follows.

Within the Gaussian {\it ansatz} as in eq.~\eqref{ansatz_2dgpe}, fluctuations are accounted for via an additional term in the energy functional~\cite{baillie2016droplet}
\begin{align}
    V_{\text{LHY}} = c \frac{\hbar^2}{m} (N a)^{5/2} \frac{1 + \frac{3}{2} \frac{ a_{dd}^2}{ a^2 }}{\sigma_\rho^3 \sigma_z^{3/2}},
\end{align}
where $c = 2^{19/2} / 75 \sqrt{5} \pi^{7 / 4} \approx 0.35$. The correspondence between the hyperspherical picture and the Gaussian {\it ansatz} to the GPE motivates the inclusion of an LHY hyperspherical potential. The Gaussian widths set the hyperradii as in eqs.~(\ref{eq:correspodance1},~\ref{eq:correspodance2}). We thus have
\begin{align}
    V_{\text{LHY}} = \frac{c}{2^{3/4}} \frac{\hbar^2}{m} (N a)^{5/2} \frac{1 + \frac{3}{2} \frac{ a_{dd}^2}{ a^2 }}{P^3 Z^{3/2}}.
\end{align}
This potential should be a good approximation in the large $N$ limit relevant to experiments, although is still not formally justified then. Deriving such a correction from first principles in the hyperspherical formalism remains an outstanding question. In this limit, the rest of the hyperspherical potential surface simplifies, and we are left with
\begin{widetext}
\begin{align}
    \bigg[ - \frac{ \hbar^2 }{ 2M } \left( \frac{ \partial^2 }{ \partial P^2 } + \frac{ \partial^2 }{ \partial Z^2 } \right)
+ \underbrace{\frac{ \hbar^2 }{ 2M }\frac{ N^2 }{ P^2 } +\frac{ \hbar^2 }{ 2M } \frac{ N^2 }{ 4 Z^2 } + \frac{1}{2} M \omega_\rho^2 P^2 + \frac{1}{2 } M \omega_z^2 Z^2 + V_c +  V_{dd} + V_{\text{LHY}}}_{V_{\text{eff}}} \bigg] F_0 = E F_0.
\label{eq:simplified_schrodinger}
\end{align}
\end{widetext}

\begin{figure}[tb]
  \centering
  \includegraphics[width = 0.48\textwidth]{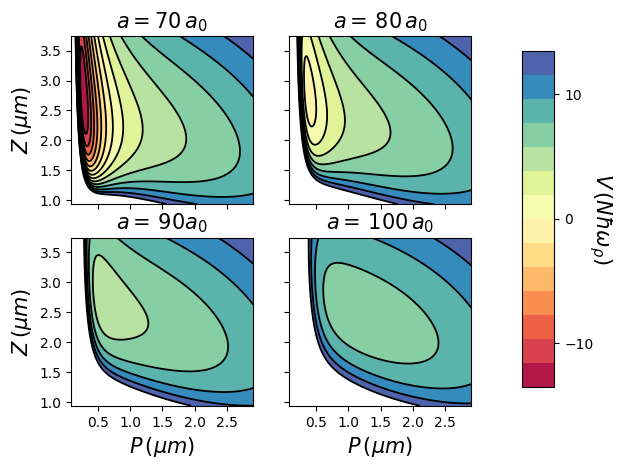}
  \caption{Potential surfaces $V_{\text{eff}}$ for $2 \times 10^4$ $^{164}$Dy atoms in a $70 \times 2\pi$~Hz spherical trap. The scattering length $a$ takes values $70,80,90$ and $100~a_0$. The droplet, given by the minimum on the far left subplots in the first row, is present for $a = 70~a_0$ and $a = 80~a_0$, but not for $a = 90~a_0$ nor $a = 100~a_0$.}
  \label{V_surfs}
\end{figure}

Using the large $N$ limit of the interaction terms in $V_\text{eff}$, we find (see appendix~\ref{app:full_hyp}) 
\begin{align}
     V_c(P,Z) &= \frac{\hbar^2 a }{2m\sqrt{\pi}}\frac{N^2}{Z P^2}
\end{align}
for the contact term where $a$ is the scattering length, and
\begin{align}
    V_{dd} &= \frac{\hbar^2 a_{dd} N^2}{4m\sqrt{\pi}} \frac{1}{Z P^2} \Bigg[ 4 + \frac{12}{\lambda^2 - 2} \nonumber \\
    &- \frac{6 \sqrt{2} \lambda^2}{(\lambda^2 -2)^{3/2}} \cot^{-1}\left( \sqrt{\frac{2}{\lambda^2 -2}}\right)  \Bigg]
\end{align}
for the dipole term. Here $\lambda = P/Z$. Although these could have been determined from the correspondence to the energy functional in the Gaussian {\it ansatz} to the GPE, appendix~\ref{app:full_hyp} gives a complete derivation of these terms within the hyperspherical formalism outside of the large $N$ limit. 

\section{Droplet to gas transition in a spherical trap}

We now apply the hyperspherical method to a collection of dipolar Dysprosium atoms held in a spherical trap. We consider system parameters similar to Ref.~\cite{wachtler2016ground,baillie2017collective} where, as the scattering length is increased, the ground state evolves smoothly from a self-bound droplet to a gaseous dipolar BEC held together by the trap potential.  In this case the ground state and collective excitation spectrum were explored within the Gaussian {\it ansatz}~\cite{wachtler2016ground} and by calculating the Bogoliubov excitation spectrum~\cite{baillie2017collective}.  Our emphasis here will be on presenting these spectra in terms of the wave functions in our linear, two-dimensional Schr\"odinger equation, and especially how these wave functions evolve across the transition.

 \subsection{Potential energy surface}
 We consider a BEC with $N = 2 \times 10^4$ $^{164}$Dy atoms in a harmonic trap with $\omega_\rho = \omega_z =70 \times 2 \pi ~\text{Hz} $. The atoms are aligned along the $z$-axis and have dipole length $a_{dd} = 131~a_0$. Figure~\ref{V_surfs} shows the resulting hyperspherical potential surface at four different scattering lengths in a spherically symmetric trap. 
 
For $a = 70~a_0$ (top left) the minimum corresponds to a droplet state that is deeply bound at $P = 0.28 ~\mu \text{m}$ and $Z = 2.73 ~\mu\text{m}$. This minimum is present without the trapping potential, and thus the state located here is a self-bound droplet. Because $Z > P$, the physical density profile of the droplet is elongated along the $z$-axis. At $a = 80 \, a_0$ (top right), the droplet is less deeply bound, although this minimum would still exist in the absence of the trap. Here the droplet is somewhat wider, with $P = 0.39 ~\mu \text{m}$ while $Z = 2.75~\mu \text{m}$, a significant change from $a = 70 \, a_0$ in the width of the condensate while the height is nearly constant.

Note that for all the potential surfaces shown, there is only one local minimum. This is not be the case in different trap geometries, which are squeezed more tightly in $z$ than $\rho$. In these pancake traps, there are two local minima which may coexist, corresponding to the pancake shaped gas and the droplet. As the scattering length changes in this scenario, the global condensate ground state abruptly changes from the droplet to the gas as one increases $a$.  

 \begin{figure*}[tb]
  \centering
  \includegraphics[width = \textwidth]{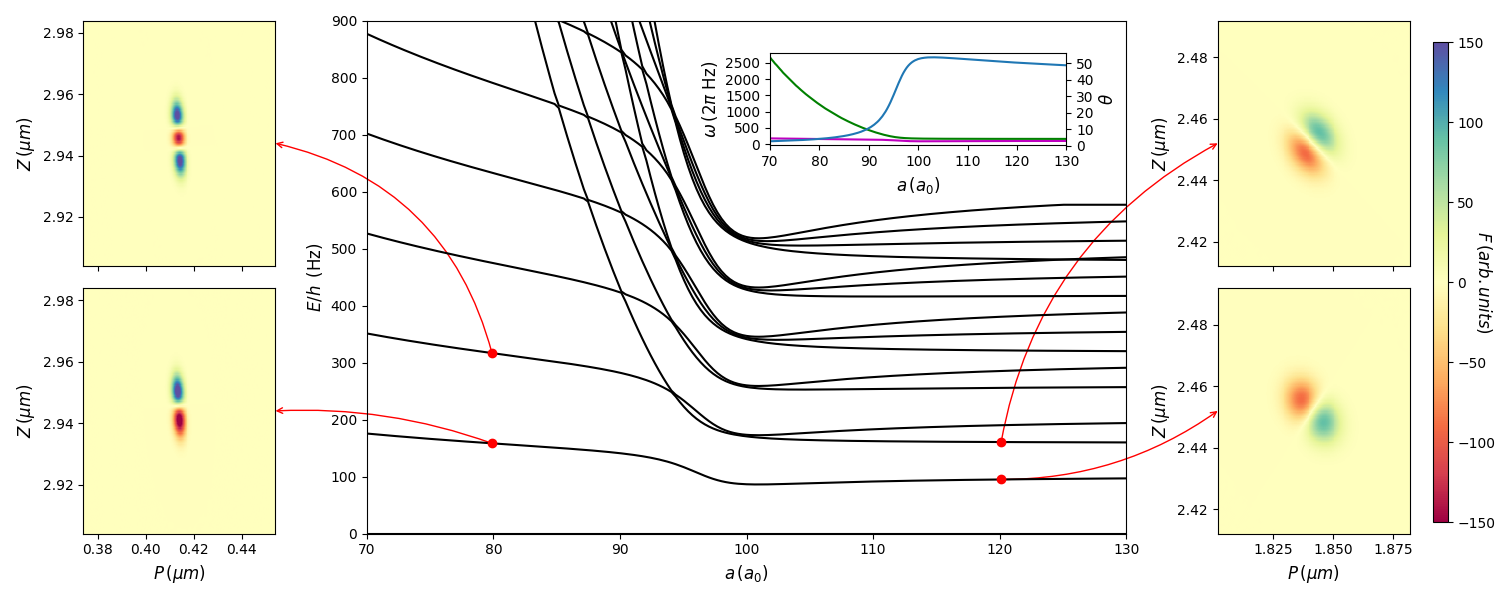}
  \caption{The spectrum of excited states relative to the ground state energy as a function of the scattering length. Here $N = 2 \times 10^4$ and $a_{dd} = 131 a_0$. The left portion of the plot shows excitations of the dipolar droplet while the right portion shows excitations of the trapped gas. A smooth transition occurs between these two regimes as the modes soften. The first two excited states are shown for two characteristic scattering length of $80 \, a_0$ and $120 \, a_0$ - on left when the BEC is in the droplet phase and on right when the BEC is in the gaseous phase. Note the different axes in the far left and far right columns, while the scale remains constant.}
  \label{fig:spectrum_states}
\end{figure*}

\subsection{Spectrum across the transition}

These potential energy surfaces correspond to the energy functional as generated by the Gaussian variational {\it ansatz} to the Gross-Pitaevskii equation.  If that functional is treated {\it as if} it were a potential energy surface~\cite{wachtler2016ground}, it can be used to assess the character of the low-lying modes by examining harmonic expansions of the functional around its minima.  By contrast, the hyperspherical approach explicitly shows the role of this potential surface, and thus can generate excited state spectra and wave functions directly from our effective 2D Schr\"odinger equation, eq.~\eqref{eq:simplified_schrodinger}, without using any harmonic approximations.

A portion of the spectrum is shown as a function of scattering length in the center panel of
Fig.~\ref{fig:spectrum_states}, which is reminiscent of the Bogoliubov spectrum presented in Fig. 3 of Ref.~\cite{baillie2017collective}.  It shares the essential feature of that figure, namely, the softening of the excited state energies as the scattering length passes from low to high.  Note that Fig. 3 of Ref.~\cite{baillie2017collective} contains all angular momentum projections $m=0,1, \dots 5$, whereas Fig.~\ref{fig:spectrum_states} of the present paper corresponds only to states with $m=0$, as considering 2 hyperradii in the K-harmonic approximation maintains cylindrical symmetry for all modes. 
 
For scattering lengths $a<90~a_0$, the spectrum consists of two sets of relatively evenly spaced levels, with very different spacings between the two sets.  At $a=90~a_0$, the lower-energy set of levels, with spacing $158$ Hz, belong to excitations along the $Z$ axis, as is verified by the sample wave functions plotted to the left of the main figure.  The higher energies, with characteristic spacings $1213$ Hz, correspond to excitations in the $P$ coordinate.  Even here, a harmonic approximation gives this higher excitation energy instead as $1225$ Hz, a $1\%$ discrepancy. The potential energy surface is nearly separable in these coordinates, as might be expected from the potentials shown in Fig.~\ref{V_surfs}.  These excitations in $P$ rapidly decline in energy as $a$ grows, since the outward pressure of a larger scattering length acts to broaden the gas in the radial direction, as also seen in Fig.~\ref{V_surfs}. 
 
As the scattering length increases beyond around $90~a_0$, the energy levels coalesce into bands with levels nearly degenerate in each band.  For $a> 100~a_0$, the levels become more evenly spaced and depend only weakly on scattering length.  In this regime, the BEC is well described by a mean-field picture, i.e. quantum fluctuations are not required for stability. Here the spectrum resembles that of  a 2D harmonic oscillator with primary excitation frequencies $E/h = 95~\text{Hz}$ and $E/h = 161~\text{Hz}$. In this case, the energy levels nearly exactly match what one would expect from the effective harmonic excitation frequencies of the hyperspherical potential for all energies shown.  

The right column of Fig.~\ref{fig:spectrum_states} shows wavefunctions for the first two excited states of the system at $a = 120~a_0$.  Here, in the gaseous state of the dipolar BEC, the symmetry of the wave functions has changed. The lowest excitation has a nodal line running from lower left to upper right.  This corresponds to an excitation where as $P$ gets large, $Z$ gets small, and vice-versa: a quadrupole mode.  In the next excited state, the nodal line runs from upper left to lower right.  In this excitation both $P$ and $Z$ grow and shrink in phase, as in a breathing mode.

We can characterize normal modes of the hyperspherical potential near its minimum.  This is similar in spirit to the modes generated by the Gaussian {\it ansatz} method~\cite{wachtler2016ground} but of course in our picture corresponds to the physical modes of an actual potential surface. The inset of Fig.~\ref{fig:spectrum_states} shows these two effective frequencies (green and magenta lines) plotted on the left-hand axis.

The normal modes of the hyperspherical potential surface describe oscillations in the $P$-$Z$ plane that occur along tilt angles that are roughly $\theta = 0$ and $\theta = \pi/2$ relative to the $Z$ axis. For a surface whose expansion near its minimum is given by
\begin{align}
    V(P,Z) = A P^2 + B Z^2 + C P Z,
    \label{eq:harmonic}
\end{align}
with $A,B,C > 0$, the tilt angle is given by
\begin{align}
    \theta = \tan^{-1} \left( \frac{C(B-A)}{(A-B)^2 + \sqrt{(A-B)^4 + (A-B)^2 C^2 }}\right)
\end{align}
for $A \neq B$. When $A=B$ the tilt angle is not well defined. In the inset to Fig.~\ref{fig:spectrum_states}, this tilt angle is shown as a blue line using the right-hand axis. 
The tilt evolves from $\theta = 4.2^\circ $ for $a = 80~a_0$, to $\theta = 50.4^\circ $ for $a=120~a_0$, thus quantifying the rotation from either $P$ or $Z$ dominated modes in the droplet, to breathing and quadrupole modes in the gaseous state.

 \begin{figure}[htb]
  \centering
  \includegraphics[width = 0.48\textwidth]{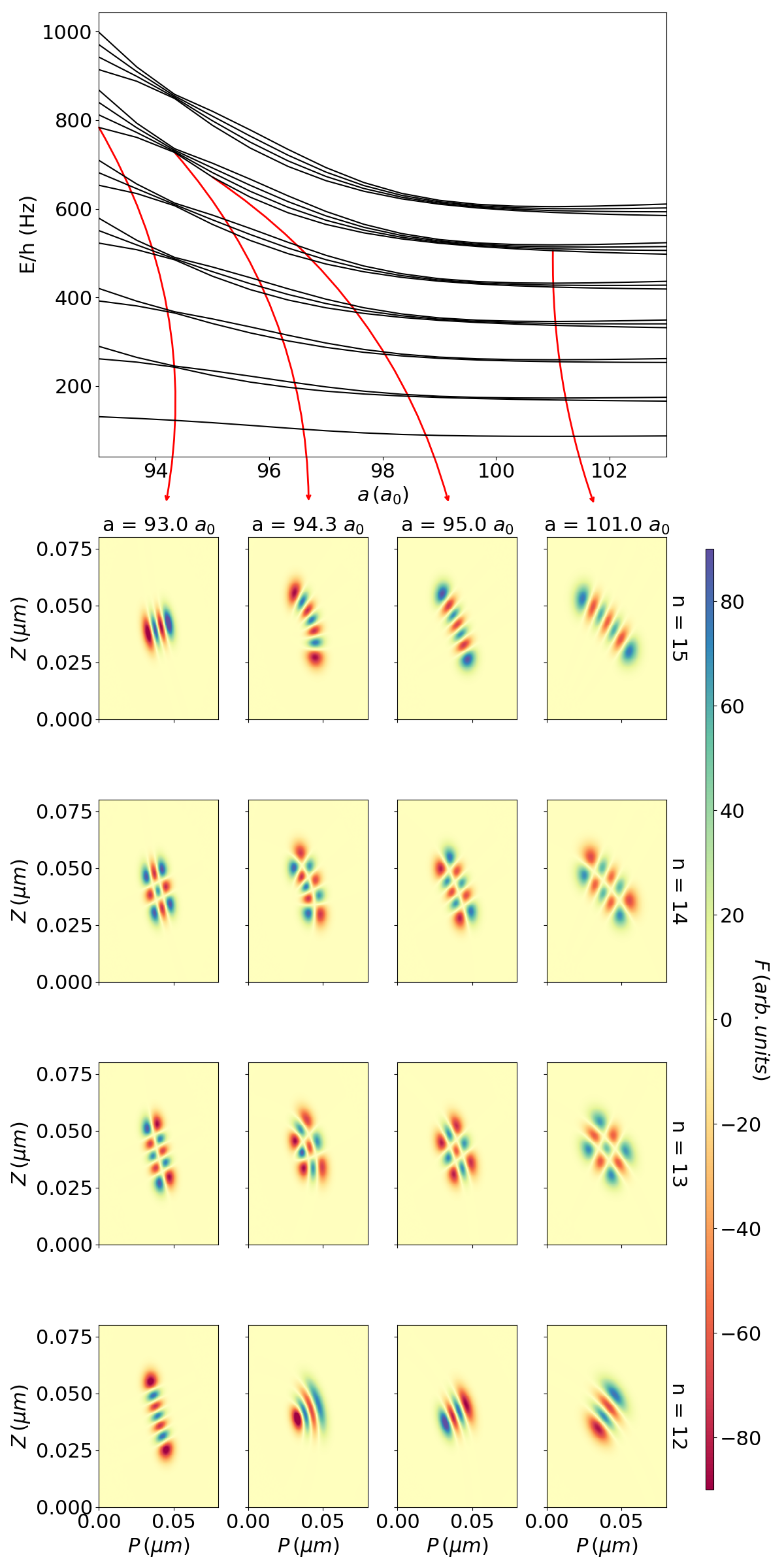}
  \caption{Spectrum and excited states while the dipolar BEC transitions between the droplet and gaseous phases, with the spectrum shown in the top panels are excited states corresponding to the red dots in the spectrum shown below. The excited states are not described by harmonic oscillator states, exhibiting curved and distorted profiles as the BEC transitions from the droplet to gaseous phase.}
  \label{cool_states}
\end{figure}

\subsection{Evolution of eigenstates}

In both of these regimes, where the dipolar BEC is clearly in one of the gaseous or droplet phases, analysis of the energy surface~\cite{wachtler2016ground} elucidates the essential properties described by the hyperspherical approach. However, in the intermediate regime, where excitation frequencies are seen to fall sharply as $a$ increases and the system transitions from the droplet to the gas, such harmonic approximations begin to fail. 
In this region, the energy levels are seen to form into bands. In each band, as $a$ gets smaller than $90~a_0$, all but one excitation, corresponding to states with at least one quantum in the radial direction, increases rapidly. Here a single energy level peels off and forms the $Z$-excitation spectrum of the droplet.

The evolution of the character of these states across the transition is best illustrated by states that are somewhat excited above the ground state.  Accordingly,
Fig.~\ref{cool_states} shows a zoomed in spectrum in this critical regime as well as states from one such band of excited states, namely the $n=12$ through $n=15$ states at four different values of $a$ in this regime. The top panel of Fig.~\ref{cool_states} shows this spectrum for the first 19 excited states. The energy levels are grouped into bands of an increasing number of states, with $\lfloor n/2 + 1 \rfloor$ states in the $n^\text{th}$ band. These bands all exhibit a crossing occurring at roughly the same value of $a$, namely $a \approx 94.3~a_0$. All the given energies levels in a band are nearly degenerate at this point. Furthermore, the energy levels again become nearly degenerate around $a=100~a_0$. The red arrows in the $6^\text{th}$ energy band indicate the energies levels shown below. 

The bottom portion of Fig.~\ref{cool_states} shows the excited states with $n=15$, $n=14$, $n=13$, and $n=12$, as labeled, corresponding to the red arrows in the above spectrum.  Each of these states is shown at 4 different values of $a$, with the first, second, third, and fourth column showing the states at $a=93 \, a_0$, $a=94.3 \, a_0$, $a= 95 \, a_0$ and $a=101 \, a_0$, respectively. For simplicity, the x and y axes both start at $0$, so the position of the excited states in the $P$-$Z$ plane is not shown. This way of plotting emphasizes comparing the shapes of the excited states, without regard to how they follow the minimum of the hyperspherical potential.

In the droplet, (left column, $a=93~a_0$), the states follow the general pattern established above.  The $n=12$ state represents an excitation principally along the elongated, $z$-axis of the droplet, while $n=15$, higher in energy, is principally a radial excitation.  Note that there is already a degree of tilt in these patterns, indicating that the hyperspherical potential energy surface is not quite separable in the $P$, $Z$ coordinates for excited states, but only approximately so. The intermediate states with $n=13,14$, containing quanta in both the $P$ and $Z$ directions. 

Past the regions of avoided crossings and into the gaseous regime, the states at $a=101 a_0$ (right-hand column) have a different character.  Here the $n=12$ state, lowest in energy, has a nodal pattern describing a breathing mode of the gas.  The higher-lying $n=15$ state shows the excitation pattern of a quadrupole.  And again, the intermediate states reveal quanta in both modes.

In the transition regime this is no longer the case.  Consider the value $a=94.3~a_0$ (second column), corresponding to the first energy level crossing seen in the transition region, where the four given states are nearly degenerate, although not exactly so. The  $n=15$ state (second column, first row), exhibits a clearly curved character. Here the state is clearly not decomposable into 2 harmonic states about any 2 straight axes. The second state shown in this column, the $n=14$ state, looks similar, with another excitation in the direction orthogonal to the curvature of the wavefunction. The state at $n=13$ has one more such quantum, while $n=12$  possesses  only excitations in this other direction. This non-separable behavior is the way the BEC negotiates its transition between the modes of the droplet and those of the gas. Even somewhat away from the degeneracy, at $a=95 a_0$, the excited states nonetheless exhibit a curved character. 

We can additionally track the states at the same excitation number between the first and second columns. Following the level crossing, the states at the same excitation number look completely different than before. However, the $n=15$ state at $a=93 \, a_0$ (first column, first row) state closely resembles the $n=12$ at $a=94.3 \, a_0$ (second column, fourth row), with some distortion. Likewise for the $n=12$ state at $a=93 \, a_0$ (first column, fourth row), and the $n=15$ state at $a=94.3 \, a_0$ (second column, first row). These two states swap spectral positions while tracking continuously changing wavefunctions. The same effect can be seen looking at states $n=13$ and $n=14$ moving between $a=93 \, a_0$ and $a=94.3 \, a_0$, where these two states swap places. The final two columns show the states before and after their second avoided crossing in the spectrum in this range of scattering length. However, here the state at a given excitation number looks much the same on either side of the crossing. As one goes to larger scattering lengths, the curved natures of these states begins to disappear. The states once again become well described by harmonic oscillator modes in two tilted directions.

 \section{Conclusion}

We applied a hyperspherical approach to a dipolar quantum gas. In order to describe dipolar droplets, we first demonstrated a general correspondence between the hyperspherical approach and the Gaussian {\it ansatz} to the GPE. This allowed us to translate the LHY correction to our 2D Schr\"odinger equation, which removes the nonlinearity from the description of dipolar BEC.  We showed excited state energies and wavefunctions of the dipolar droplet. Especially in the phase transition regime, one must take into account the full potential where it is not well approximated by a harmonic oscillator in both directions.  

The hyperspherical approach thus presents an intriguing middle ground between the full Bogoliubov excited spectrum, and the intuitive Gaussian variational {\it ansatz} method. The hyperspherical approach preserves the appealing intuition of the latter, but affords its extension to higher excited collective modes that lie beyond a simple expression as separable modes in an effective harmonic oscillator. Further extensions of the approach could account for the tunneling of barely-bound droplets into the gaseous state, much as the macroscopic tunneling of BECs with attractive interactions was studied~\cite{bohn1998effective}.  Additionally, the hyperspherical approach could be employed to follow dynamics of dipolar gases via the usual expansion of a linear system into energy eigenstates,  providing an alternative view to the numerical evolution of the nonlinear EGPE.
\\

This material is based upon work supported by the National Science Foundation under Grant Number PHY 1734006.

\bibliography{dipoles}

\begin{thebibliography}{41}%
\makeatletter
\providecommand \@ifxundefined [1]{%
 \@ifx{#1\undefined}
}%
\providecommand \@ifnum [1]{%
 \ifnum #1\expandafter \@firstoftwo
 \else \expandafter \@secondoftwo
 \fi
}%
\providecommand \@ifx [1]{%
 \ifx #1\expandafter \@firstoftwo
 \else \expandafter \@secondoftwo
 \fi
}%
\providecommand \natexlab [1]{#1}%
\providecommand \enquote  [1]{``#1''}%
\providecommand \bibnamefont  [1]{#1}%
\providecommand \bibfnamefont [1]{#1}%
\providecommand \citenamefont [1]{#1}%
\providecommand \href@noop [0]{\@secondoftwo}%
\providecommand \href [0]{\begingroup \@sanitize@url \@href}%
\providecommand \@href[1]{\@@startlink{#1}\@@href}%
\providecommand \@@href[1]{\endgroup#1\@@endlink}%
\providecommand \@sanitize@url [0]{\catcode `\\12\catcode `\$12\catcode
  `\&12\catcode `\#12\catcode `\^12\catcode `\_12\catcode `\%12\relax}%
\providecommand \@@startlink[1]{}%
\providecommand \@@endlink[0]{}%
\providecommand \url  [0]{\begingroup\@sanitize@url \@url }%
\providecommand \@url [1]{\endgroup\@href {#1}{\urlprefix }}%
\providecommand \urlprefix  [0]{URL }%
\providecommand \Eprint [0]{\href }%
\providecommand \doibase [0]{http://dx.doi.org/}%
\providecommand \selectlanguage [0]{\@gobble}%
\providecommand \bibinfo  [0]{\@secondoftwo}%
\providecommand \bibfield  [0]{\@secondoftwo}%
\providecommand \translation [1]{[#1]}%
\providecommand \BibitemOpen [0]{}%
\providecommand \bibitemStop [0]{}%
\providecommand \bibitemNoStop [0]{.\EOS\space}%
\providecommand \EOS [0]{\spacefactor3000\relax}%
\providecommand \BibitemShut  [1]{\csname bibitem#1\endcsname}%
\let\auto@bib@innerbib\@empty
\bibitem [{\citenamefont {Ferrier-Barbut}\ \emph {et~al.}(2016)\citenamefont
  {Ferrier-Barbut}, \citenamefont {Kadau}, \citenamefont {Schmitt},
  \citenamefont {Wenzel},\ and\ \citenamefont {Pfau}}]{ferrier2016observation}%
  \BibitemOpen
  \bibfield  {author} {\bibinfo {author} {\bibfnamefont {I.}~\bibnamefont
  {Ferrier-Barbut}}, \bibinfo {author} {\bibfnamefont {H.}~\bibnamefont
  {Kadau}}, \bibinfo {author} {\bibfnamefont {M.}~\bibnamefont {Schmitt}},
  \bibinfo {author} {\bibfnamefont {M.}~\bibnamefont {Wenzel}}, \ and\ \bibinfo
  {author} {\bibfnamefont {T.}~\bibnamefont {Pfau}},\ }\bibfield  {title}
  {\enquote {\bibinfo {title} {Observation of quantum droplets in a strongly
  dipolar bose gas},}\ }\href@noop {} {\bibfield  {journal} {\bibinfo
  {journal} {Phys. Rev. Lett.}\ }\textbf {\bibinfo {volume} {116}},\ \bibinfo
  {pages} {215301} (\bibinfo {year} {2016})}\BibitemShut {NoStop}%
\bibitem [{\citenamefont {Kadau}\ \emph {et~al.}(2016)\citenamefont {Kadau},
  \citenamefont {Schmitt}, \citenamefont {Wenzel}, \citenamefont {Wink},
  \citenamefont {Maier}, \citenamefont {Ferrier-Barbut},\ and\ \citenamefont
  {Pfau}}]{kadau2016observing}%
  \BibitemOpen
  \bibfield  {author} {\bibinfo {author} {\bibfnamefont {H.}~\bibnamefont
  {Kadau}}, \bibinfo {author} {\bibfnamefont {M.}~\bibnamefont {Schmitt}},
  \bibinfo {author} {\bibfnamefont {M.}~\bibnamefont {Wenzel}}, \bibinfo
  {author} {\bibfnamefont {C.}~\bibnamefont {Wink}}, \bibinfo {author}
  {\bibfnamefont {T.}~\bibnamefont {Maier}}, \bibinfo {author} {\bibfnamefont
  {I.}~\bibnamefont {Ferrier-Barbut}}, \ and\ \bibinfo {author} {\bibfnamefont
  {T.}~\bibnamefont {Pfau}},\ }\bibfield  {title} {\enquote {\bibinfo {title}
  {Observing the rosensweig instability of a quantum ferrofluid},}\ }\href@noop
  {} {\bibfield  {journal} {\bibinfo  {journal} {Nature}\ }\textbf {\bibinfo
  {volume} {530}},\ \bibinfo {pages} {194--197} (\bibinfo {year}
  {2016})}\BibitemShut {NoStop}%
\bibitem [{\citenamefont {Chomaz}\ \emph {et~al.}(2016)\citenamefont {Chomaz},
  \citenamefont {Baier}, \citenamefont {Petter}, \citenamefont {Mark},
  \citenamefont {W{\"a}chtler}, \citenamefont {Santos},\ and\ \citenamefont
  {Ferlaino}}]{chomaz2016quantum}%
  \BibitemOpen
  \bibfield  {author} {\bibinfo {author} {\bibfnamefont {L}~\bibnamefont
  {Chomaz}}, \bibinfo {author} {\bibfnamefont {S}~\bibnamefont {Baier}},
  \bibinfo {author} {\bibfnamefont {D}~\bibnamefont {Petter}}, \bibinfo
  {author} {\bibfnamefont {MJ}~\bibnamefont {Mark}}, \bibinfo {author}
  {\bibfnamefont {F}~\bibnamefont {W{\"a}chtler}}, \bibinfo {author}
  {\bibfnamefont {Luis}\ \bibnamefont {Santos}}, \ and\ \bibinfo {author}
  {\bibfnamefont {F}~\bibnamefont {Ferlaino}},\ }\bibfield  {title} {\enquote
  {\bibinfo {title} {Quantum-fluctuation-driven crossover from a dilute
  bose-einstein condensate to a macrodroplet in a dipolar quantum fluid},}\
  }\href@noop {} {\bibfield  {journal} {\bibinfo  {journal} {Physical Review
  X}\ }\textbf {\bibinfo {volume} {6}},\ \bibinfo {pages} {041039} (\bibinfo
  {year} {2016})}\BibitemShut {NoStop}%
\bibitem [{\citenamefont {Schmitt}\ \emph {et~al.}(2016)\citenamefont
  {Schmitt}, \citenamefont {B{\"o}ttcher}, \citenamefont {Ferrier-Barbut},\
  and\ \citenamefont {Pfau}}]{schmitt2016self}%
  \BibitemOpen
  \bibfield  {author} {\bibinfo {author} {\bibfnamefont {M.}~\bibnamefont
  {Schmitt}, \bibfnamefont {M.and~Wenzel}}, \bibinfo {author} {\bibfnamefont
  {F.}~\bibnamefont {B{\"o}ttcher}}, \bibinfo {author} {\bibfnamefont
  {I.}~\bibnamefont {Ferrier-Barbut}}, \ and\ \bibinfo {author} {\bibfnamefont
  {T.}~\bibnamefont {Pfau}},\ }\bibfield  {title} {\enquote {\bibinfo {title}
  {Self-bound droplets of a dilute magnetic quantum liquid},}\ }\href@noop {}
  {\bibfield  {journal} {\bibinfo  {journal} {Nature}\ }\textbf {\bibinfo
  {volume} {539}},\ \bibinfo {pages} {259--262} (\bibinfo {year}
  {2016})}\BibitemShut {NoStop}%
\bibitem [{\citenamefont {Lima}\ and\ \citenamefont
  {Pelster}(2011)}]{lima2011quantum}%
  \BibitemOpen
  \bibfield  {author} {\bibinfo {author} {\bibfnamefont {A.~R.~P.}\
  \bibnamefont {Lima}}\ and\ \bibinfo {author} {\bibfnamefont {A.}~\bibnamefont
  {Pelster}},\ }\bibfield  {title} {\enquote {\bibinfo {title} {Quantum
  fluctuations in dipolar bose gases},}\ }\href@noop {} {\bibfield  {journal}
  {\bibinfo  {journal} {Phys. Rev. A}\ }\textbf {\bibinfo {volume} {84}},\
  \bibinfo {pages} {041604} (\bibinfo {year} {2011})}\BibitemShut {NoStop}%
\bibitem [{\citenamefont {Baillie}\ \emph {et~al.}(2016)\citenamefont
  {Baillie}, \citenamefont {Wilson}, \citenamefont {Bisset},\ and\
  \citenamefont {Blakie}}]{baillie2016droplet}%
  \BibitemOpen
  \bibfield  {author} {\bibinfo {author} {\bibfnamefont {D.}~\bibnamefont
  {Baillie}}, \bibinfo {author} {\bibfnamefont {R.~M.}\ \bibnamefont {Wilson}},
  \bibinfo {author} {\bibfnamefont {R.~N.}\ \bibnamefont {Bisset}}, \ and\
  \bibinfo {author} {\bibfnamefont {P.~B.}\ \bibnamefont {Blakie}},\ }\bibfield
   {title} {\enquote {\bibinfo {title} {Self-bound dipolar droplet: A localized
  matter wave in free space},}\ }\href {\doibase 10.1103/PhysRevA.94.021602}
  {\bibfield  {journal} {\bibinfo  {journal} {Phys. Rev. A}\ }\textbf {\bibinfo
  {volume} {94}},\ \bibinfo {pages} {021602} (\bibinfo {year}
  {2016})}\BibitemShut {NoStop}%
\bibitem [{\citenamefont {Bisset}\ \emph
  {et~al.}(2016{\natexlab{a}})\citenamefont {Bisset}, \citenamefont {Wilson},
  \citenamefont {Baillie},\ and\ \citenamefont {Blakie}}]{bisset2016ground}%
  \BibitemOpen
  \bibfield  {author} {\bibinfo {author} {\bibfnamefont {R.~N.}\ \bibnamefont
  {Bisset}}, \bibinfo {author} {\bibfnamefont {R.~M.}\ \bibnamefont {Wilson}},
  \bibinfo {author} {\bibfnamefont {D.}~\bibnamefont {Baillie}}, \ and\
  \bibinfo {author} {\bibfnamefont {P.~B.}\ \bibnamefont {Blakie}},\ }\bibfield
   {title} {\enquote {\bibinfo {title} {Ground-state phase diagram of a dipolar
  condensate with quantum fluctuations},}\ }\href@noop {} {\bibfield  {journal}
  {\bibinfo  {journal} {Phys. Rev. A}\ }\textbf {\bibinfo {volume} {94}},\
  \bibinfo {pages} {033619} (\bibinfo {year} {2016}{\natexlab{a}})}\BibitemShut
  {NoStop}%
\bibitem [{\citenamefont {Baillie}\ \emph {et~al.}(2017)\citenamefont
  {Baillie}, \citenamefont {Wilson},\ and\ \citenamefont
  {Blakie}}]{baillie2017collective}%
  \BibitemOpen
  \bibfield  {author} {\bibinfo {author} {\bibfnamefont {D.}~\bibnamefont
  {Baillie}}, \bibinfo {author} {\bibfnamefont {R.~M.}\ \bibnamefont {Wilson}},
  \ and\ \bibinfo {author} {\bibfnamefont {P.~B.}\ \bibnamefont {Blakie}},\
  }\bibfield  {title} {\enquote {\bibinfo {title} {Collective excitations of
  self-bound droplets of a dipolar quantum fluid},}\ }\href@noop {} {\bibfield
  {journal} {\bibinfo  {journal} {Phys. Rev. Lett.}\ }\textbf {\bibinfo
  {volume} {119}},\ \bibinfo {pages} {255302} (\bibinfo {year}
  {2017})}\BibitemShut {NoStop}%
\bibitem [{\citenamefont {W\"achtler}\ and\ \citenamefont
  {Santos}(2016)}]{wachtler2016ground}%
  \BibitemOpen
  \bibfield  {author} {\bibinfo {author} {\bibfnamefont {F.}~\bibnamefont
  {W\"achtler}}\ and\ \bibinfo {author} {\bibfnamefont {L.}~\bibnamefont
  {Santos}},\ }\bibfield  {title} {\enquote {\bibinfo {title} {Ground-state
  properties and elementary excitations of quantum droplets in dipolar
  bose-einstein condensates},}\ }\href {\doibase 10.1103/PhysRevA.94.043618}
  {\bibfield  {journal} {\bibinfo  {journal} {Phys. Rev. A}\ }\textbf {\bibinfo
  {volume} {94}},\ \bibinfo {pages} {043618} (\bibinfo {year}
  {2016})}\BibitemShut {NoStop}%
\bibitem [{\citenamefont {Bohn}\ \emph {et~al.}(1998)\citenamefont {Bohn},
  \citenamefont {Esry},\ and\ \citenamefont {Greene}}]{bohn1998effective}%
  \BibitemOpen
  \bibfield  {author} {\bibinfo {author} {\bibfnamefont {J.~L.}\ \bibnamefont
  {Bohn}}, \bibinfo {author} {\bibfnamefont {B.~D.}\ \bibnamefont {Esry}}, \
  and\ \bibinfo {author} {\bibfnamefont {C.~H.}\ \bibnamefont {Greene}},\
  }\bibfield  {title} {\enquote {\bibinfo {title} {Effective potentials for
  dilute bose-einstein condensates},}\ }\href@noop {} {\bibfield  {journal}
  {\bibinfo  {journal} {Phys. Rev. A}\ }\textbf {\bibinfo {volume} {58}},\
  \bibinfo {pages} {584} (\bibinfo {year} {1998})}\BibitemShut {NoStop}%
\bibitem [{\citenamefont {Watson}\ and\ \citenamefont
  {McKinney}(1999)}]{watson1999improved}%
  \BibitemOpen
  \bibfield  {author} {\bibinfo {author} {\bibfnamefont {D.~K.}\ \bibnamefont
  {Watson}}\ and\ \bibinfo {author} {\bibfnamefont {B.~A.}\ \bibnamefont
  {McKinney}},\ }\bibfield  {title} {\enquote {\bibinfo {title} {Improved
  large-n limit for bose-einstein condensates from perturbation theory},}\
  }\href@noop {} {\bibfield  {journal} {\bibinfo  {journal} {Phys. Rev. A}\
  }\textbf {\bibinfo {volume} {59}},\ \bibinfo {pages} {4091} (\bibinfo {year}
  {1999})}\BibitemShut {NoStop}%
\bibitem [{\citenamefont {Biswas}\ \emph {et~al.}(2010)\citenamefont {Biswas},
  \citenamefont {Das}, \citenamefont {Salasnich},\ and\ \citenamefont
  {Chakrabarti}}]{biswas2010stability}%
  \BibitemOpen
  \bibfield  {author} {\bibinfo {author} {\bibfnamefont {A.}~\bibnamefont
  {Biswas}}, \bibinfo {author} {\bibfnamefont {T.~K.}\ \bibnamefont {Das}},
  \bibinfo {author} {\bibfnamefont {L.}~\bibnamefont {Salasnich}}, \ and\
  \bibinfo {author} {\bibfnamefont {B.}~\bibnamefont {Chakrabarti}},\
  }\bibfield  {title} {\enquote {\bibinfo {title} {Stability of an attractive
  bosonic cloud with van der waals interaction},}\ }\href@noop {} {\bibfield
  {journal} {\bibinfo  {journal} {Phys. Rev. A}\ }\textbf {\bibinfo {volume}
  {82}},\ \bibinfo {pages} {043607} (\bibinfo {year} {2010})}\BibitemShut
  {NoStop}%
\bibitem [{\citenamefont {Haldar}\ \emph {et~al.}(2014)\citenamefont {Haldar},
  \citenamefont {Chakrabarti}, \citenamefont {Bhattacharyya},\ and\
  \citenamefont {Das}}]{haldar2014condensate}%
  \BibitemOpen
  \bibfield  {author} {\bibinfo {author} {\bibfnamefont {S.~K.}\ \bibnamefont
  {Haldar}}, \bibinfo {author} {\bibfnamefont {B.}~\bibnamefont {Chakrabarti}},
  \bibinfo {author} {\bibfnamefont {S.}~\bibnamefont {Bhattacharyya}}, \ and\
  \bibinfo {author} {\bibfnamefont {T.~K.}\ \bibnamefont {Das}},\ }\bibfield
  {title} {\enquote {\bibinfo {title} {Condensate fraction and critical
  temperature of interacting bose gas in anharmonic trap},}\ }\href@noop {}
  {\bibfield  {journal} {\bibinfo  {journal} {The European Physical Journal D}\
  }\textbf {\bibinfo {volume} {68}},\ \bibinfo {pages} {1--10} (\bibinfo {year}
  {2014})}\BibitemShut {NoStop}%
\bibitem [{\citenamefont {Sogo}\ \emph {et~al.}(2004)\citenamefont {Sogo},
  \citenamefont {S{\o}rensen}, \citenamefont {Jensen},\ and\ \citenamefont
  {Fedorov}}]{sogo2004stability}%
  \BibitemOpen
  \bibfield  {author} {\bibinfo {author} {\bibfnamefont {T.}~\bibnamefont
  {Sogo}}, \bibinfo {author} {\bibfnamefont {O.}~\bibnamefont {S{\o}rensen}},
  \bibinfo {author} {\bibfnamefont {A.~S.}\ \bibnamefont {Jensen}}, \ and\
  \bibinfo {author} {\bibfnamefont {D.~V.}\ \bibnamefont {Fedorov}},\
  }\bibfield  {title} {\enquote {\bibinfo {title} {Stability and structure of
  two coupled boson systems in an external field},}\ }\href@noop {} {\bibfield
  {journal} {\bibinfo  {journal} {Phys. Rev. A}\ }\textbf {\bibinfo {volume}
  {69}},\ \bibinfo {pages} {062504} (\bibinfo {year} {2004})}\BibitemShut
  {NoStop}%
\bibitem [{\citenamefont {S{\o}rensen}\ \emph {et~al.}(2002)\citenamefont
  {S{\o}rensen}, \citenamefont {Fedorov},\ and\ \citenamefont
  {Jensen}}]{sorensen2002two}%
  \BibitemOpen
  \bibfield  {author} {\bibinfo {author} {\bibfnamefont {O.}~\bibnamefont
  {S{\o}rensen}}, \bibinfo {author} {\bibfnamefont {D.~V.}\ \bibnamefont
  {Fedorov}}, \ and\ \bibinfo {author} {\bibfnamefont {A.~S.}\ \bibnamefont
  {Jensen}},\ }\bibfield  {title} {\enquote {\bibinfo {title} {Two-body
  correlations in n-body boson systems},}\ }\href@noop {} {\bibfield  {journal}
  {\bibinfo  {journal} {Phys. Rev. A}\ }\textbf {\bibinfo {volume} {66}},\
  \bibinfo {pages} {032507} (\bibinfo {year} {2002})}\BibitemShut {NoStop}%
\bibitem [{\citenamefont {S{\o}rensen}\ \emph
  {et~al.}(2003{\natexlab{a}})\citenamefont {S{\o}rensen}, \citenamefont
  {Fedorov},\ and\ \citenamefont {Jensen}}]{sorensen2003correlated}%
  \BibitemOpen
  \bibfield  {author} {\bibinfo {author} {\bibfnamefont {O.}~\bibnamefont
  {S{\o}rensen}}, \bibinfo {author} {\bibfnamefont {D.~V.}\ \bibnamefont
  {Fedorov}}, \ and\ \bibinfo {author} {\bibfnamefont {A.~S.}\ \bibnamefont
  {Jensen}},\ }\bibfield  {title} {\enquote {\bibinfo {title} {Correlated
  n-boson systems for arbitrary scattering length},}\ }\href@noop {} {\bibfield
   {journal} {\bibinfo  {journal} {Phys. Rev. A}\ }\textbf {\bibinfo {volume}
  {68}},\ \bibinfo {pages} {063618} (\bibinfo {year}
  {2003}{\natexlab{a}})}\BibitemShut {NoStop}%
\bibitem [{\citenamefont {S{\o}rensen}\ \emph
  {et~al.}(2003{\natexlab{b}})\citenamefont {S{\o}rensen}, \citenamefont
  {Fedorov},\ and\ \citenamefont {Jensen}}]{sorensen2003structure}%
  \BibitemOpen
  \bibfield  {author} {\bibinfo {author} {\bibfnamefont {O.}~\bibnamefont
  {S{\o}rensen}}, \bibinfo {author} {\bibfnamefont {D.~V.}\ \bibnamefont
  {Fedorov}}, \ and\ \bibinfo {author} {\bibfnamefont {A.~S.}\ \bibnamefont
  {Jensen}},\ }\bibfield  {title} {\enquote {\bibinfo {title} {Structure of
  boson systems beyond the mean field},}\ }\href@noop {} {\bibfield  {journal}
  {\bibinfo  {journal} {Journal Phys. B}\ }\textbf {\bibinfo {volume} {37}},\
  \bibinfo {pages} {93} (\bibinfo {year} {2003}{\natexlab{b}})}\BibitemShut
  {NoStop}%
\bibitem [{\citenamefont {Th{\o}gersen}\ \emph {et~al.}(2007)\citenamefont
  {Th{\o}gersen}, \citenamefont {Fedorov},\ and\ \citenamefont
  {Jensen}}]{thogersen2007trapped}%
  \BibitemOpen
  \bibfield  {author} {\bibinfo {author} {\bibfnamefont {M.}~\bibnamefont
  {Th{\o}gersen}}, \bibinfo {author} {\bibfnamefont {D.~V.}\ \bibnamefont
  {Fedorov}}, \ and\ \bibinfo {author} {\bibfnamefont {A.~S.}\ \bibnamefont
  {Jensen}},\ }\bibfield  {title} {\enquote {\bibinfo {title} {Trapped bose
  gases with large positive scattering length},}\ }\href@noop {} {\bibfield
  {journal} {\bibinfo  {journal} {EPL (Euro. Phys. Lett.}\ }\textbf {\bibinfo
  {volume} {79}},\ \bibinfo {pages} {40002} (\bibinfo {year}
  {2007})}\BibitemShut {NoStop}%
\bibitem [{\citenamefont {Lekala}\ \emph {et~al.}(2014)\citenamefont {Lekala},
  \citenamefont {Chakrabarti}, \citenamefont {Rampho}, \citenamefont {Das},
  \citenamefont {Sofianos},\ and\ \citenamefont {Adam}}]{lekala2014behavior}%
  \BibitemOpen
  \bibfield  {author} {\bibinfo {author} {\bibfnamefont {M.~L.}\ \bibnamefont
  {Lekala}}, \bibinfo {author} {\bibfnamefont {B.}~\bibnamefont {Chakrabarti}},
  \bibinfo {author} {\bibfnamefont {G.~J.}\ \bibnamefont {Rampho}}, \bibinfo
  {author} {\bibfnamefont {T.~K.}\ \bibnamefont {Das}}, \bibinfo {author}
  {\bibfnamefont {S.~A.}\ \bibnamefont {Sofianos}}, \ and\ \bibinfo {author}
  {\bibfnamefont {R.~M.}\ \bibnamefont {Adam}},\ }\bibfield  {title} {\enquote
  {\bibinfo {title} {Behavior of trapped ultracold dilute bose gases at large
  scattering length near a feshbach resonance},}\ }\href@noop {} {\bibfield
  {journal} {\bibinfo  {journal} {Phys. Rev. A}\ }\textbf {\bibinfo {volume}
  {89}},\ \bibinfo {pages} {023624} (\bibinfo {year} {2014})}\BibitemShut
  {NoStop}%
\bibitem [{\citenamefont {Ding}\ and\ \citenamefont
  {Greene}(2017)}]{ding2017renormalized}%
  \BibitemOpen
  \bibfield  {author} {\bibinfo {author} {\bibfnamefont {Y.}~\bibnamefont
  {Ding}}\ and\ \bibinfo {author} {\bibfnamefont {C.~H.}\ \bibnamefont
  {Greene}},\ }\bibfield  {title} {\enquote {\bibinfo {title} {Renormalized
  contact interaction in degenerate unitary bose gases},}\ }\href@noop {}
  {\bibfield  {journal} {\bibinfo  {journal} {Phys. Rev. A}\ }\textbf {\bibinfo
  {volume} {95}},\ \bibinfo {pages} {053602} (\bibinfo {year}
  {2017})}\BibitemShut {NoStop}%
\bibitem [{\citenamefont {Sze}\ \emph {et~al.}(2018)\citenamefont {Sze},
  \citenamefont {Sykes}, \citenamefont {Blume},\ and\ \citenamefont
  {Bohn}}]{sze2018hyperspherical}%
  \BibitemOpen
  \bibfield  {author} {\bibinfo {author} {\bibfnamefont {M.~W.~C.}\
  \bibnamefont {Sze}}, \bibinfo {author} {\bibfnamefont {A.~G.}\ \bibnamefont
  {Sykes}}, \bibinfo {author} {\bibfnamefont {D.}~\bibnamefont {Blume}}, \ and\
  \bibinfo {author} {\bibfnamefont {J.~L.}\ \bibnamefont {Bohn}},\ }\bibfield
  {title} {\enquote {\bibinfo {title} {Hyperspherical lowest-order
  constrained-variational approximation to resonant bose-einstein
  condensates},}\ }\href@noop {} {\bibfield  {journal} {\bibinfo  {journal}
  {Phys. Rev. A}\ }\textbf {\bibinfo {volume} {97}},\ \bibinfo {pages} {033608}
  (\bibinfo {year} {2018})}\BibitemShut {NoStop}%
\bibitem [{\citenamefont {Das}\ and\ \citenamefont
  {Chakrabarti}(2004)}]{das2004potential}%
  \BibitemOpen
  \bibfield  {author} {\bibinfo {author} {\bibfnamefont {T.~K.}\ \bibnamefont
  {Das}}\ and\ \bibinfo {author} {\bibfnamefont {B.}~\bibnamefont
  {Chakrabarti}},\ }\bibfield  {title} {\enquote {\bibinfo {title} {Potential
  harmonics expansion method for trapped interacting bosons: Inclusion of
  two-body correlation},}\ }\href@noop {} {\bibfield  {journal} {\bibinfo
  {journal} {Phys. Rev. A}\ }\textbf {\bibinfo {volume} {70}},\ \bibinfo
  {pages} {063601} (\bibinfo {year} {2004})}\BibitemShut {NoStop}%
\bibitem [{\citenamefont {Das}\ \emph {et~al.}(2007)\citenamefont {Das},
  \citenamefont {Canuto}, \citenamefont {Kundu},\ and\ \citenamefont
  {Chakrabarti}}]{das2007behavior}%
  \BibitemOpen
  \bibfield  {author} {\bibinfo {author} {\bibfnamefont {T.~K.}\ \bibnamefont
  {Das}}, \bibinfo {author} {\bibfnamefont {S.}~\bibnamefont {Canuto}},
  \bibinfo {author} {\bibfnamefont {A.}~\bibnamefont {Kundu}}, \ and\ \bibinfo
  {author} {\bibfnamefont {B.}~\bibnamefont {Chakrabarti}},\ }\bibfield
  {title} {\enquote {\bibinfo {title} {Behavior of a bose-einstein condensate
  containing a large number of atoms interacting through a finite-range
  interatomic interaction},}\ }\href@noop {} {\bibfield  {journal} {\bibinfo
  {journal} {Phys. Rev. A}\ }\textbf {\bibinfo {volume} {75}},\ \bibinfo
  {pages} {042705} (\bibinfo {year} {2007})}\BibitemShut {NoStop}%
\bibitem [{\citenamefont {Chakrabarti}\ and\ \citenamefont
  {Das}(2008)}]{chakrabarti2008shape}%
  \BibitemOpen
  \bibfield  {author} {\bibinfo {author} {\bibfnamefont {B.}~\bibnamefont
  {Chakrabarti}}\ and\ \bibinfo {author} {\bibfnamefont {T.~K.}\ \bibnamefont
  {Das}},\ }\bibfield  {title} {\enquote {\bibinfo {title} {Shape-independent
  approximation for bose-einstein condensates interacting through a van der
  waals potential},}\ }\href@noop {} {\bibfield  {journal} {\bibinfo  {journal}
  {Phys. Rev. A}\ }\textbf {\bibinfo {volume} {78}},\ \bibinfo {pages} {063608}
  (\bibinfo {year} {2008})}\BibitemShut {NoStop}%
\bibitem [{\citenamefont {van~der Hart}(2000)}]{van2000collapse}%
  \BibitemOpen
  \bibfield  {author} {\bibinfo {author} {\bibfnamefont {H.~W.}\ \bibnamefont
  {van~der Hart}},\ }\bibfield  {title} {\enquote {\bibinfo {title} {Collapse
  versus growth for a bose-einstein condensate with attractive interactions},}\
  }\href@noop {} {\bibfield  {journal} {\bibinfo  {journal} {Phy. Rev. A}\
  }\textbf {\bibinfo {volume} {62}},\ \bibinfo {pages} {013601} (\bibinfo
  {year} {2000})}\BibitemShut {NoStop}%
\bibitem [{\citenamefont {Sogo}\ \emph {et~al.}(2005)\citenamefont {Sogo},
  \citenamefont {Fedorov},\ and\ \citenamefont {Jensen}}]{sogo2005coherent}%
  \BibitemOpen
  \bibfield  {author} {\bibinfo {author} {\bibfnamefont {T.}~\bibnamefont
  {Sogo}}, \bibinfo {author} {\bibfnamefont {D.~V.}\ \bibnamefont {Fedorov}}, \
  and\ \bibinfo {author} {\bibfnamefont {A.~S.}\ \bibnamefont {Jensen}},\
  }\bibfield  {title} {\enquote {\bibinfo {title} {Coherent atom--molecule
  oscillations with hyperspherical coordinates},}\ }\href@noop {} {\bibfield
  {journal} {\bibinfo  {journal} {J. Phys. B}\ }\textbf {\bibinfo {volume}
  {38}},\ \bibinfo {pages} {2979} (\bibinfo {year} {2005})}\BibitemShut
  {NoStop}%
\bibitem [{\citenamefont {Liu}\ \emph {et~al.}(2007)\citenamefont {Liu},
  \citenamefont {Morishita},\ and\ \citenamefont {Watanabe}}]{liu2007time}%
  \BibitemOpen
  \bibfield  {author} {\bibinfo {author} {\bibfnamefont {C.-N.}\ \bibnamefont
  {Liu}}, \bibinfo {author} {\bibfnamefont {.}~\bibnamefont {Morishita}}, \
  and\ \bibinfo {author} {\bibfnamefont {S.}~\bibnamefont {Watanabe}},\
  }\bibfield  {title} {\enquote {\bibinfo {title} {Time-dependent
  hyperspherical studies for a two-dimensional attractive bose-einstein
  condensate},}\ }\href@noop {} {\bibfield  {journal} {\bibinfo  {journal}
  {Phys. Rev. A}\ }\textbf {\bibinfo {volume} {75}},\ \bibinfo {pages} {023604}
  (\bibinfo {year} {2007})}\BibitemShut {NoStop}%
\bibitem [{\citenamefont {Lee}\ and\ \citenamefont
  {Greene}(2021)}]{PhysRevA.103.023325}%
  \BibitemOpen
  \bibfield  {author} {\bibinfo {author} {\bibfnamefont {H.}~\bibnamefont
  {Lee}}\ and\ \bibinfo {author} {\bibfnamefont {C.~H.}\ \bibnamefont
  {Greene}},\ }\bibfield  {title} {\enquote {\bibinfo {title} {Orbital
  variational adiabatic hyperspherical method applied to bose-einstein
  condensates},}\ }\href {\doibase 10.1103/PhysRevA.103.023325} {\bibfield
  {journal} {\bibinfo  {journal} {Phys. Rev. A}\ }\textbf {\bibinfo {volume}
  {103}},\ \bibinfo {pages} {023325} (\bibinfo {year} {2021})}\BibitemShut
  {NoStop}%
\bibitem [{\citenamefont {Rittenhouse}\ \emph {et~al.}(2006)\citenamefont
  {Rittenhouse}, \citenamefont {Cavagnero}, \citenamefont {von Stecher},\ and\
  \citenamefont {Greene}}]{rittenhouse2006hyperspherical}%
  \BibitemOpen
  \bibfield  {author} {\bibinfo {author} {\bibfnamefont {S.~T.}\ \bibnamefont
  {Rittenhouse}}, \bibinfo {author} {\bibfnamefont {M.~J.}\ \bibnamefont
  {Cavagnero}}, \bibinfo {author} {\bibfnamefont {J}~\bibnamefont {von
  Stecher}}, \ and\ \bibinfo {author} {\bibfnamefont {C.~H.}\ \bibnamefont
  {Greene}},\ }\bibfield  {title} {\enquote {\bibinfo {title} {Hyperspherical
  description of the degenerate fermi gas: s-wave interactions},}\ }\href@noop
  {} {\bibfield  {journal} {\bibinfo  {journal} {Phys. Rev. A}\ }\textbf
  {\bibinfo {volume} {74}},\ \bibinfo {pages} {053624} (\bibinfo {year}
  {2006})}\BibitemShut {NoStop}%
\bibitem [{\citenamefont {Rittenhouse}\ and\ \citenamefont
  {Greene}(2008)}]{rittenhouse2008degenerate}%
  \BibitemOpen
  \bibfield  {author} {\bibinfo {author} {\bibfnamefont {S.~T.}\ \bibnamefont
  {Rittenhouse}}\ and\ \bibinfo {author} {\bibfnamefont {C.~H.}\ \bibnamefont
  {Greene}},\ }\bibfield  {title} {\enquote {\bibinfo {title} {The degenerate
  fermi gas with density-dependent interactions in the large-n limit under the
  k-harmonic approximation},}\ }\href@noop {} {\bibfield  {journal} {\bibinfo
  {journal} {J. Phys. B: At. Mol. Opt. Phys.}\ }\textbf {\bibinfo {volume}
  {41}},\ \bibinfo {pages} {205302} (\bibinfo {year} {2008})}\BibitemShut
  {NoStop}%
\bibitem [{\citenamefont {Rittenhouse}\ \emph {et~al.}(2009)\citenamefont
  {Rittenhouse}, \citenamefont {Cavagnero},\ and\ \citenamefont
  {Greene}}]{rittenhouse2009collective}%
  \BibitemOpen
  \bibfield  {author} {\bibinfo {author} {\bibfnamefont {S.~T.}\ \bibnamefont
  {Rittenhouse}}, \bibinfo {author} {\bibfnamefont {M.~J.}\ \bibnamefont
  {Cavagnero}}, \ and\ \bibinfo {author} {\bibfnamefont {C.~H.}\ \bibnamefont
  {Greene}},\ }\bibfield  {title} {\enquote {\bibinfo {title} {Collective
  coordinate description of anisotropically trapped degenerate fermi gases},}\
  }\href@noop {} {\bibfield  {journal} {\bibinfo  {journal} {J. Phys. Chem. A}\
  }\textbf {\bibinfo {volume} {113}},\ \bibinfo {pages} {15016--15023}
  (\bibinfo {year} {2009})}\BibitemShut {NoStop}%
\bibitem [{\citenamefont {Kushibe}\ \emph {et~al.}(2004)\citenamefont
  {Kushibe}, \citenamefont {Mutou}, \citenamefont {Morishita}, \citenamefont
  {Watanabe},\ and\ \citenamefont {Matsuzawa}}]{kushibe2004aspects}%
  \BibitemOpen
  \bibfield  {author} {\bibinfo {author} {\bibfnamefont {D.}~\bibnamefont
  {Kushibe}}, \bibinfo {author} {\bibfnamefont {M.}~\bibnamefont {Mutou}},
  \bibinfo {author} {\bibfnamefont {T.}~\bibnamefont {Morishita}}, \bibinfo
  {author} {\bibfnamefont {S.}~\bibnamefont {Watanabe}}, \ and\ \bibinfo
  {author} {\bibfnamefont {M.}~\bibnamefont {Matsuzawa}},\ }\bibfield  {title}
  {\enquote {\bibinfo {title} {Aspects of hyperspherical adiabaticity in an
  atomic-gas bose-einstein condensate},}\ }\href@noop {} {\bibfield  {journal}
  {\bibinfo  {journal} {Phys. Rev. A}\ }\textbf {\bibinfo {volume} {70}},\
  \bibinfo {pages} {063617} (\bibinfo {year} {2004})}\BibitemShut {NoStop}%
\bibitem [{\citenamefont {Blakie}\ \emph {et~al.}(2020)\citenamefont {Blakie},
  \citenamefont {Baillie},\ and\ \citenamefont {Pal}}]{blakie2020variational}%
  \BibitemOpen
  \bibfield  {author} {\bibinfo {author} {\bibfnamefont {P.~B.}\ \bibnamefont
  {Blakie}}, \bibinfo {author} {\bibfnamefont {D.}~\bibnamefont {Baillie}}, \
  and\ \bibinfo {author} {\bibfnamefont {S.}~\bibnamefont {Pal}},\ }\bibfield
  {title} {\enquote {\bibinfo {title} {Variational theory for the ground state
  and collective excitations of an elongated dipolar condensate},}\ }\href@noop
  {} {\bibfield  {journal} {\bibinfo  {journal} {Commun. Theor. Phys.}\
  }\textbf {\bibinfo {volume} {72}},\ \bibinfo {pages} {085501} (\bibinfo
  {year} {2020})}\BibitemShut {NoStop}%
\bibitem [{\citenamefont {Hu}\ and\ \citenamefont
  {Liu}(2020)}]{hu2020collective}%
  \BibitemOpen
  \bibfield  {author} {\bibinfo {author} {\bibfnamefont {H.}~\bibnamefont
  {Hu}}\ and\ \bibinfo {author} {\bibfnamefont {X.-J.}\ \bibnamefont {Liu}},\
  }\bibfield  {title} {\enquote {\bibinfo {title} {Collective excitations of a
  spherical ultradilute quantum droplet},}\ }\href@noop {} {\bibfield
  {journal} {\bibinfo  {journal} {Phys. Rev. A}\ }\textbf {\bibinfo {volume}
  {102}},\ \bibinfo {pages} {053303} (\bibinfo {year} {2020})}\BibitemShut
  {NoStop}%
\bibitem [{\citenamefont {Huang}\ and\ \citenamefont {Yang}(1957)}]{LHY1}%
  \BibitemOpen
  \bibfield  {author} {\bibinfo {author} {\bibfnamefont {K.}~\bibnamefont
  {Huang}}\ and\ \bibinfo {author} {\bibfnamefont {Chen~N.}\ \bibnamefont
  {Yang}},\ }\bibfield  {title} {\enquote {\bibinfo {title} {Quantum-mechanical
  many-body problem with hard-sphere interaction},}\ }\href@noop {} {\bibfield
  {journal} {\bibinfo  {journal} {Phys. Rev.}\ }\textbf {\bibinfo {volume}
  {105}},\ \bibinfo {pages} {767} (\bibinfo {year} {1957})}\BibitemShut
  {NoStop}%
\bibitem [{\citenamefont {Lee}\ \emph {et~al.}(1957)\citenamefont {Lee},
  \citenamefont {Huang},\ and\ \citenamefont {Yang}}]{LHY2}%
  \BibitemOpen
  \bibfield  {author} {\bibinfo {author} {\bibfnamefont {T.~D.}\ \bibnamefont
  {Lee}}, \bibinfo {author} {\bibfnamefont {K.}~\bibnamefont {Huang}}, \ and\
  \bibinfo {author} {\bibfnamefont {C.~N.}\ \bibnamefont {Yang}},\ }\bibfield
  {title} {\enquote {\bibinfo {title} {Eigenvalues and eigenfunctions of a bose
  system of hard spheres and its low-temperature properties},}\ }\href@noop {}
  {\bibfield  {journal} {\bibinfo  {journal} {Phys. Rev.}\ }\textbf {\bibinfo
  {volume} {106}},\ \bibinfo {pages} {1135} (\bibinfo {year}
  {1957})}\BibitemShut {NoStop}%
\bibitem [{\citenamefont {Smirnov}\ and\ \citenamefont
  {Shitikova}(1977)}]{Smirnov77}%
  \BibitemOpen
  \bibfield  {author} {\bibinfo {author} {\bibfnamefont {Y.~F.}\ \bibnamefont
  {Smirnov}}\ and\ \bibinfo {author} {\bibfnamefont {K.~V.}\ \bibnamefont
  {Shitikova}},\ }\bibfield  {title} {\enquote {\bibinfo {title} {Method of k
  harmonics and the shell model},}\ }\href@noop {} {\bibfield  {journal}
  {\bibinfo  {journal} {Sov. J. Particles Nucl.}\ }\textbf {\bibinfo {volume}
  {8}} (\bibinfo {year} {1977})}\BibitemShut {NoStop}%
\bibitem [{\citenamefont {Avery}\ \emph {et~al.}(1997)\citenamefont {Avery},
  \citenamefont {Bian}, \citenamefont {Loeser},\ and\ \citenamefont
  {Antonsen}}]{avery1997fourier}%
  \BibitemOpen
  \bibfield  {author} {\bibinfo {author} {\bibfnamefont {J.}~\bibnamefont
  {Avery}}, \bibinfo {author} {\bibfnamefont {W.}~\bibnamefont {Bian}},
  \bibinfo {author} {\bibfnamefont {J.}~\bibnamefont {Loeser}}, \ and\ \bibinfo
  {author} {\bibfnamefont {F.}~\bibnamefont {Antonsen}},\ }\bibfield  {title}
  {\enquote {\bibinfo {title} {Fourier transform approach to potential
  harmonics},}\ }\href@noop {} {\bibfield  {journal} {\bibinfo  {journal}
  {International journal of quantum chemistry}\ }\textbf {\bibinfo {volume}
  {63}},\ \bibinfo {pages} {5--14} (\bibinfo {year} {1997})}\BibitemShut
  {NoStop}%
\bibitem [{\citenamefont {Dalfovo}\ \emph {et~al.}(1999)\citenamefont
  {Dalfovo}, \citenamefont {Giorgini}, \citenamefont {Pitaevskii},\ and\
  \citenamefont {Stringari}}]{dalfovo1999theory}%
  \BibitemOpen
  \bibfield  {author} {\bibinfo {author} {\bibfnamefont {F.}~\bibnamefont
  {Dalfovo}}, \bibinfo {author} {\bibfnamefont {S.}~\bibnamefont {Giorgini}},
  \bibinfo {author} {\bibfnamefont {L.~P.}\ \bibnamefont {Pitaevskii}}, \ and\
  \bibinfo {author} {\bibfnamefont {S.}~\bibnamefont {Stringari}},\ }\bibfield
  {title} {\enquote {\bibinfo {title} {Theory of bose-einstein condensation in
  trapped gases},}\ }\href@noop {} {\bibfield  {journal} {\bibinfo  {journal}
  {Rev. Mod. Phys.}\ }\textbf {\bibinfo {volume} {71}},\ \bibinfo {pages} {463}
  (\bibinfo {year} {1999})}\BibitemShut {NoStop}%
\bibitem [{\citenamefont {Bisset}\ \emph
  {et~al.}(2016{\natexlab{b}})\citenamefont {Bisset}, \citenamefont {Wilson},
  \citenamefont {Baillie},\ and\ \citenamefont {Blakie}}]{PhysRevA.94.033619}%
  \BibitemOpen
  \bibfield  {author} {\bibinfo {author} {\bibfnamefont {R.~N.}\ \bibnamefont
  {Bisset}}, \bibinfo {author} {\bibfnamefont {R.~M.}\ \bibnamefont {Wilson}},
  \bibinfo {author} {\bibfnamefont {D.}~\bibnamefont {Baillie}}, \ and\
  \bibinfo {author} {\bibfnamefont {P.~B.}\ \bibnamefont {Blakie}},\ }\bibfield
   {title} {\enquote {\bibinfo {title} {Ground-state phase diagram of a dipolar
  condensate with quantum fluctuations},}\ }\href {\doibase
  10.1103/PhysRevA.94.033619} {\bibfield  {journal} {\bibinfo  {journal} {Phys.
  Rev. A}\ }\textbf {\bibinfo {volume} {94}},\ \bibinfo {pages} {033619}
  (\bibinfo {year} {2016}{\natexlab{b}})}\BibitemShut {NoStop}%
\bibitem [{\citenamefont {Gradshteyn}\ and\ \citenamefont
  {Ryzhik}(1994)}]{BigRussianBook}%
  \BibitemOpen
  \bibfield  {author} {\bibinfo {author} {\bibfnamefont {I.~S.}\ \bibnamefont
  {Gradshteyn}}\ and\ \bibinfo {author} {\bibfnamefont {I.~M.}\ \bibnamefont
  {Ryzhik}},\ }\href@noop {} {\emph {\bibinfo {title} {Table of integrals,
  series, and products}}}\ (\bibinfo  {publisher} {Academic Press},\ \bibinfo
  {year} {1994})\BibitemShut {NoStop}%
\end{thebibliography}%

\appendix
\section{Direct Calculation of Hyperspherical Integrals}
\label{app:full_hyp}

Here we obtain expressions for the hyperspherical potential surface for arbitrary $N$. We  compute the effective potentials for both contact and dipolar interactions, and then verify that these do indeed reduce to simpler forms that match the Gaussian {\it ansatz} to the GPE in the large $N$ limit. For a two-body contact potential of the form
\begin{align}
   V_c(\mathbf{r_{ij}}) =\frac{4 \pi \hbar^2 a}{m} \delta(\mathbf{r_{ij}}), 
\end{align}
we need to find
\begin{align}
     V_c(P,Z) &= \sum_{i<j} \int d\Omega Y_{00}^*(\Omega) V_c({\bf r}_{ij}) Y_{00}(\Omega) \\
     &= \frac{ N(N-1) }{ 2 } \int d\Omega Y_{00}^*(\Omega) V_c({\bf r}_{12}) Y_{00}(\Omega).
\end{align}
Note that $V_c(P,Z)$ refers to the effective hyperspherical potential while $ V_c(\mathbf{r_{ij}})$ refers to the actual two-body potential.
We can evaluate this directly, as the delta function vastly simplifies eq.~\eqref{exact_2d_int}. We find that
\begin{align}
V_c &=\frac{\hbar^2 a (N-1)}{\sqrt{2\pi N}m}\frac{1}{Z P^2}\frac{\Gamma(N/2)(N-1)}{\Gamma(N/2 - 1/2)}, \label{true_Vc} \\
&\approx \frac{\hbar^2 a }{2m\sqrt{\pi}}\frac{N^2}{Z P^2}.
\label{approx_Vc}
\end{align}
Here eq.~\eqref{approx_Vc} was obtained by using the large $N$ limit of the Gamma functions. This approximate expression matches the term obtained in the Gaussian {\it ansatz} to the GPE~\cite{baillie2016droplet}.

For the dipolar potential, we have
\begin{align}
V_{dd}({\bf r}) =  \frac{3\hbar^2}{m} \frac{a_{dd}}{r^3} (1-3\cos^2\theta).
\end{align}
Here the dipole length is defined as $a_{dd} \equiv m\mu_0\mu^2/12\pi\hbar^2$, and $\theta$ is the angle between the $\bf{r}$ and the polarization axis. We can find a 1D integral expression for $V_{dd}(P,Z)$, which simplifies in the large $N$ limit.

We work in momentum space because it will later allow us to deal with the $1/r^3$ singularity in the dipole-dipole potential. We define the following unit vectors~\cite{avery1997fourier}
        
\begin{align}
    \hat{u}_P &= \frac{1}{P\sqrt{N}}(x_1,y_1,x_2,y_2,\ldots,x_N,y_N) \\
    \hat{u}_Z &= \frac{1}{Z\sqrt{N}}(z_1,z_2,\ldots,z_N) \\
    \hat{w}_P &= \frac{1}{\sqrt{2} k \lvert \sin \theta_k \rvert}(k_x,k_y,-k_x,-k_y,0,\ldots,0) \\
    \hat{w}_Z &= \frac{1}{\sqrt{2} k \lvert \cos \theta_k \rvert}(k_z,-k_z,0,\ldots,0)
\end{align}
With $k$ and $\theta_k$ having the usual definitions for writing $(k_x,k_y,k_z)$ in spherical polar coordinates. Notice that $\hat{u}$ depends only on position coordinates and $\hat{w}$ depends only on momentum coordinates. Using these unit vectors, we can embed the plane wave in the larger hyperspheres~\cite{avery1997fourier}
\begin{widetext}
\begin{align}
    e^{i \vec{k} \cdot \vec{r}_{12}} &= e^{i \sqrt{2N} k P \lvert \sin \theta_k \rvert \hat{w}_P \cdot \hat{u}_P} e^{i \sqrt{2N} k Z \lvert \cos \theta_k \rvert \hat{w}_Z \cdot \hat{u}_Z} \\
    &= \frac{(2N -2)!!(N-2)!!}{Y_{\small{00}}^2} \sum_{\lambda_P,~\mu_P} i^{\lambda_P} j_{\lambda_P}^{2N} \left( \sqrt{2N} k P \lvert \sin \theta_k \rvert \right)Y_{\lambda_P\mu_P}^* \left( \hat{w}_P \right) Y_{\lambda_P\mu_P} \left( \hat{u}_P \right) \nonumber \\ &\times \sum_{\lambda_Z,~\mu_Z} i^{\lambda_Z} j_{\lambda_Z}^{N} \left( \sqrt{2N} k Z \lvert \cos \theta_k \rvert \right)Y_{\lambda_Z\mu_Z}^* \left( \hat{w}_Z \right) Y_{\lambda_Z\mu_Z} \left( \hat{u}_Z \right)
\end{align}
\end{widetext}
We embedded the first term in the $2N$ dimensional $P$ space, and the second term in the $N$ dimension $Z$ space. $j^d$ are the hyperspherical Bessel functions in $d$ dimensions. We also use the 3 dimensional Fourier transform of $V_{dd}(\vec{r}_{12})$,  ${\stackrel{\sim}{V_{dd}}}(\vec{k}) =  \frac{\hbar^2 a_{dd}}{m}\sqrt{\frac{2}{\pi}} \left( 3 \cos^2 \theta_k -1 \right) $. Now
\begin{align}
    V_{dd}(P,Z) &= \frac{N(N-1)}{2} \int d\Omega~Y_{00} V(\vec{r}_{12}) Y_{00}, \nonumber \\
    &= \frac{N(N-1)}{2}\frac{\hbar^2 a_{dd}}{2\pi^2 m} (2N -2)!!(N-2)!! \nonumber \\
    &\times \int d^3\vec{k}(3\cos^2 \theta_k - 1 ) {j_{0}^{2N}}(\alpha k){j_{0}^{N}}(\beta k).
\end{align}

Here $\alpha = \sqrt{2N} P \lvert \sin \theta_k \rvert $ and $\beta = \sqrt{2N} Z \lvert \cos \theta_k \rvert $.  The hyperspherical Bessel Functions can we written in terms of the standard spherical Bessel Functions, giving
\begin{align}
    V_{dd}(P,Z) &= N(N-1) \frac{\hbar^2 a_{dd}}{\pi^2 m} \Gamma(N)\Gamma(N/2) 2^{3N/2 -4} \nonumber \\
    &\times \underbrace{\int d^3\vec{k}(3\cos^2 \theta_k - 1) \frac{J_{N-1}(\alpha k)J_{N/2 -1}(\beta k)}{(\alpha k)^{N-1} (\beta k)^{N/2 - 1}}}_{I}
\end{align}
Then
\begin{align*}
    I = \int d\hat{k} \frac{3\cos^2 \theta_k - 1}{\alpha^{N-1}\beta^{N/2 - 1}} \underbrace{\int_0^\infty k^2 dk \frac{J_{N-1}(\alpha k) J_{N/2 -1}(\beta k)}{k^{3N/2 -2}}}_{I_k}.
\end{align*}
The integral $I_k$ needs to be broken up into two cases to be done analytically: one when $\alpha > \beta$ and one when $\alpha < \beta$. Since the integrand does not diverge when $\alpha=\beta$, we can safely split the integral up into two components. Using~\cite{BigRussianBook}, we get

\begin{align}
\small
      I_k = 
   \frac{\Gamma(3/2)}{2^{3N/2 - 4}} \begin{cases}
     \frac{\alpha^{N-1} \beta^{N/2 -4}}{\Gamma((N-3)/2)\Gamma(N)}F\left( \frac{3}{2},\frac{5-N}{2};N;x^2 \right) & x > 1 \\ \\
     \frac{\alpha^{N-4} \beta^{N/2 -1}}{\Gamma((2N-3)/2)\Gamma(N/2)}F \left( \frac{3}{2},\frac{5-2N}{2};\frac{N}{2};y^2 \right) & y > 1 . \\
    \end{cases}
\end{align}

Here we have defined $x \equiv \alpha/\beta$, $y \equiv 1/x$, while $F$ denotes the ordinary (Gaussian) hypergeometric functions. In order to proceed further here, we turn to the large N limit of these hypergeometric functions. From the definitions of these functions, we can write
\begin{align*}
    F\left( \frac{3}{2},\frac{5-N}{2};N;x^2 \right) &= \sum_{k=0}^\infty \frac{\left( \frac{3}{2}\right)_k \left( \frac{5-N}{2} \right)_k} {\left(N \right)_k} \frac{ x^{2k}}{k!},  \\
    &\approx  \sum_{k=0}^\infty \left( \frac{3}{2}\right)_k \frac{\left( -\frac{x^2}{2} \right)^{k}}{k!}. \\
     &\approx \left( 1 + x^2 \right)^{-3/2} .
\end{align*}
Here $(z)_k$ gives the rising factorial. Between the first and second lines we have simplified ratios to their limiting values in the large $N$ limit. This sum then matches the Taylor expansion of the function given in the third line. Likewise, 
\begin{align*}
F\left( \frac{3}{2},\frac{5-2N}{2};\frac{N}{2};y^2 \right) \approx \left( 1+2 y^2 \right)^{-3/2} .
\end{align*}
Since there is cylindrical symmetry, as well as mirror symmetry through the $x$-$y$ plane, we can then rewrite the angular integral as 
\begin{align*}
    I = 4\pi \int_0^{\pi/2} \sin \theta_k d\theta_k \frac{3\cos^2 \theta_k - 1}{\alpha^{N-1}\beta^{N/2 - 1}} I_k
\end{align*}
Now define  $\theta_= \equiv \cot Z/ P$. Note that $0 \leq \theta_= \leq \pi/2$, and at $\theta_=$ we have $\alpha = \beta$. For $0 \leq \theta < \theta_=$ we see that $\alpha < \beta$ and for $\pi/2 \geq \theta > \theta_=$ we see that $\beta > \alpha$. This will let us use our two cases for $I_k$.
\begin{align} 
I &=  4 \pi \int_0^{\theta_=} \sin \theta_k d\theta_k \frac{3\cos^2 \theta_k - 1}{\alpha^{N-1}\beta^{N/2 - 1}} I_k, \nonumber \\
&+ 4 \pi \int_{\theta_=}^{\pi/2} \sin \theta_k d\theta_k \frac{3\cos^2 \theta_k - 1}{\alpha^{N-1}\beta^{N/2 - 1}} I_k \\
&\approx \frac{4\pi\Gamma(3/2)}{2^{3N/2 -4}(2N)^{3/2}}\frac{1}{Z P^2} \bigg[ \frac{1}{\Gamma((N-3)/2)\Gamma(N)} \nonumber \\
&\times \int_0^1 dx \frac{x(2P^2 - Z^2 x^2)}{P^2 + Z^2x^2}\left( 1 + x^2/2 \right)^{-3/2}\nonumber\\
&+ \frac{1}{\Gamma((2N-3)/2)\Gamma(N/2)}\int_0^1 dy \frac{2P^2 y^2 - Z^2}{P^2 y^2+Z^2} \left( 1+ 2y^2 \right)^{-3/2} \bigg] .
\end{align}
Where used our definitions for $x,y$ and then plugged in the asymptotic form of the hypergeometric functions. Finally, $x$ and $y$ have become dummy variables in the integral, and we can join these into a single integral in $x$. These equations will be useful in calculating the contact potential. So we can now write down $V_{dd}$ in the large $N$ limit as:
\begin{align}
V_{dd} &\approx \frac{\hbar^2 a_{dd} N^2}{4m\sqrt{\pi}} \frac{1}{Z P^2} \Bigg[ \int_0^1 dx \frac{x(2P^2 - Z^2 x^2)}{P^2 + Z^2x^2}\left( 1 + x^2/2 \right)^{-3/2} \nonumber \\
&+ 2\sqrt{2}\int_0^1 dy \frac{2P^2 y^2 - Z^2}{P^2 y^2+Z^2} \left( 1+ 2y^2 \right)^{-3/2} \Bigg].
\end{align}
Where we use the limiting form of the ratios of $\Gamma$ functions, which converge quite rapidly. These integrals have a closed form expression, when we express this in terms of the aspect ratio, $\lambda = P/Z$. Using~\cite{BigRussianBook}, we arrive at
\begin{align}
V_{dd} &\approx \frac{\hbar^2 a_{dd} N^2}{4m\sqrt{\pi}} \frac{1}{Z P^2} \Bigg[ 4 + \frac{12}{\lambda^2 - 2} \nonumber \\
&- \frac{6 \sqrt{2} \lambda^2}{(\lambda^2 -2.)^{3/2}} \cot^{-1}\left( \sqrt{\frac{2}{\lambda^2 -2}}\right)  \Bigg]
\end{align}
Note that there is no issue with divergence at $\lambda = \sqrt{2}$, as the diverging terms of this expression cancel there. This matches the large $N$ limit of the relevant term in the Gaussian {\it ansatz} to the GPE~\cite{baillie2016droplet}.

\end{document}